\documentclass[12pt]{article}
\usepackage{feynmf}
\usepackage{epsf}
\def\la{\mathrel{\mathpalette\fun <}}
\def\ga{\mathrel{\mathpalette\fun >}}
\def\fun#1#2{\lower3.6pt\vbox{\baselineskip0pt\lineskip.9pt
\ialign{$\mathsurround=0pt#1\hfil##\hfil$\crcr#2\crcr\sim\crcr}}}
\begin{document}
\title{Field Theory and the Standard Model}
\author{V.Novikov \\
ITEP, Moscow}
\date{}

\maketitle

\begin{abstract}

The following 5 lectures are devoted to key ideas in field theory and in the
Standard Model.

\end{abstract}

\newpage

Lecture I. {\bf Quantum Field Theory. Bird's-eye-view.}

\vspace{5mm}

{\bf Introductory remarks.}

\vspace{3mm}

  Quantum Field Theory (QFT) is the working language in the community of
high energy physicists. It is not the esoteric theory accessible to the small
group of experts, the basic ideas of QFT have to be familiar to all members
of the community.There are number of excellent textbooks on QFT. The
list of the recommended books can be found in ref [1]. As a rule they are
rather lengthy, the average size of the standard textbook is of the order of 800 pages. The only way to understand QFT is to take
one of these books and to spend one year or more to study it. It is hard way,
but nobody knows the better one. So is life!

  The goal of these lectures is not to provide any systematic introduction to
the subject. It is impossible to do in five lectures. The goal is to remind
the students what they actually had studied a few years ago. Just the basic
concepts, notions and relations of QFT without long derivations and boring
formalism.

\vspace{5mm}

{\bf 1.1. Particles and Fields.}

\vspace{3mm}

 In the Classical Physics the particles and fields are  very
different dynamical systems. Particles are particles and fields are fields.
There is no way to confuse these notions. This is everyday wisdom.

 The system of particles has finite number of degrees of freedom N. To describe
 this dynamical system one have to
introduce the general coordinates  $q_i(t)$
 ($i=1,2,..N$)  and their time derivatives  $\dot{q}_i(t)$ or conjugated
momenta $p_i(t)$.  Either we study the bounded motion or the scattering
processes at any time we can say how many degrees of freedom the system has.
 Even when we observe the decay of the system we are sure that outgoing
 particles were bounded inside the initial system before the decay. This is
 evident.

Field theory is a theory of the system with infinite
number of degrees of freedom. To describe the electromagnetic fields we have 
to know four-potential $A_{\mu}$ at every space point $x$. Maxwell equations 
govern the evolution of the field in time.

The Quantum Mechanics (QM)
 of \underline{nonrelativistic} particles was developed
in 1925-26. In QM dynamical system with $N$ degrees of freedom is described
by wave function
$$
\Psi(q,t) = \Psi(q_1, ..., q_N; t)
\eqno(1.1)
$$
that satisfies the wave equation
$$
i \frac{\partial}{\partial t} \Psi(q, t) = H(p,q) \Psi(q,t)
\eqno(1.2)
$$
where $H(p,q)$ is the Hamiltonian and $p$ is the operator of momentum: $p =
-i \partial/\partial q$. The number of degrees of freedom $N$ was supposed to
be fixed exactly like in Classical Mechanics.

The first quantization of electromagnetic fields had been done at the same
time in 1926 by Born, Heisenberg and Jordan in their second paper on QM. They
represented radiation electromagnetic field as an infinite set of harmonic
oscillators and quantized these oscillators. They found that excitations of
the oscillators behave like a free massless particles -- photons. The number
of photons was not fixed. They were created and destroyed by charged
particles. Quantized theory of electromagnetic field became a theory of
particles -- photons. Photons were not "bounded" inside charged particles,
they were created from "nothing" by scattered charged particles. The physical
idea of photons was introduced by Einstein twenty years before this paper,
but the formal quantization of field showed that quantized field is
equivalent to the system of particles that can be created and destroyed.

For some time physicists tried to find a relativistic version of wave
equation (1.2) for the particles at high energy.

The first such equation was written in 1926 by Klein and Gordon for spin 0
relativistic particle
$$
-\partial_{\mu} \partial_{\mu}\Phi(x) = m^2\Phi(x)
\eqno(1.3)
$$
where $\partial_{\mu} =\partial/\partial x_{\mu}$ and $\Phi(x)$ is a complex
function of $x=(t, \vec{x})$, $m$ is a mass of particle.

Dirac pointed out that eq. (1.3) and the function
$\Phi(x)$ can't be
interpreted as a wave equation and wave function. In 1928 he suggested his own
relativistic equation for spin $1/2$ particles:
$$
(i\gamma_{\mu}\partial_{\mu} -m)\Psi(x) = 0
\eqno(1.4)
$$
Here $\Psi$ is a column with 4 complex components (4-spinor) and
$\gamma_{\mu}$ are $4\times 4$ matrices.

The troubles with interpretation of eq. (1.4) as the one-particle
relativistic wave equation are not so evident as for the case of eq. (1.3).
But the truth is that for any relativistic processes the single particle
description breaks down. Any relativistic system has infinite numbers of
degrees of freedom. More energy we pump into the system, more degrees of
freedom can be excited. For example any scattering process in QED can be
accompanied by creation of additional $e^+ e^-$ pairs. These pairs are not
hidden inside initial particles, they are created during the scattering
process. The natural description of  relativistic physics is quantum field
theory. So it is wrong to divide world on particles and fields. We have to use
the quantum field theory for \underline{everything}.

From this point of view both the Klein-Gordon and Dirac equations are not
relativistic wave equations. They are field equations for scalar and spinor
fields. These fields have to be quantize. The lowest excitations of these
quantum fields behave like  massive particles with spin $0$ and $1/2$
respectively.

The QFT is the right language for dealing with particle physics. This
language is not unique. For example string theory also pretends to describe
particles in low energy limit. We have also to note that one can
construct "diagrammatica" (i.e. the set of rules for calculation of
amplitudes in perturbation theory) without any reference to QFT.

\vspace{5mm}

{\bf 1.2. Quantization and  the Fock space.}

\vspace{3mm}

Consider a field theory for a free scalar particle:
$$
\Phi(x) = \Phi(\vec{x},t)
\eqno(1.5)
$$
In the "Classical Theory" $\Phi(x)$ is a real function of space-time point
$x_{\mu} =(t, \vec{x})$ with the \underline{lagrangian density} ${\cal
L}(\Phi, \partial_{\mu} \Phi)$
$$
{\cal L}(\Phi, \partial_{\mu}\Phi) =
\frac{1}{2}\{\partial_{\mu}\Phi\partial_{\mu}\Phi - m^2\Phi^2 \}
\eqno(1.6)
$$
where $\partial_{\mu}\Phi = \frac{\partial}{\partial x_{\mu}}\Phi$ and $m$ is
the mass of the particle.

The \underline{action} $S$ is given by
$$
S = \int d^4 x {\cal L}(\Phi, \partial_{\mu}\Phi)
\eqno(1.7)
$$
The hamiltonian density is constructed
according to the rules of hamiltonian dynamics
$$
{\cal H} = \pi \frac{\delta{\cal L}}{S\dot{\Phi}} -{\cal L} = \frac{1}{2} \{
\pi^2 +(\nabla\Phi)^2 +m^2\Phi^2\}
\eqno(1.8)
$$
where
$$
\pi = \frac{\delta{\cal L}}{\delta\Phi} = \dot{\Phi} = \partial_0\Phi
$$

We use the natural units where $c\equiv 1$ and $\hbar\equiv 1$. So the action
is dimensionless
$$
[S] = m^0 \;\; ,
$$
and for other quantities we get
$$
[E] = [p] =m
$$
$$
[x] = m^{-1}
\eqno(1.9)
$$
$$
[{\cal L}] = [{\cal H}] = m^4
$$
$$
[\phi] = m
$$

\vspace{3mm}

{\it Exercise:} prove the dimension rule eq. (1.9)

Equations of motion are derived from Hamilton variational principle
$$
\left\{
\begin{array}{ll}
\delta S = 0 \\
\partial_{\mu}\left(\frac{\delta {\cal L}}{\delta\partial_{\mu}\Phi}
\right) = \frac{\delta {\cal L}}{\delta\Phi}
\end{array}
\right.
\eqno(1.10)
$$
Eiler-Lagrange equations (1.10) for the density eq. (1.6) coincides with
Klein-Gordon equation
$$
(\partial^2 +m^2)\Phi =0
\eqno(1.11)
$$
Consider the plane wave anzatz for the solution of eq. (1.11)
$$
\Phi_{\vec{p}}(x,t) = a(t)e^{i\vec{p}\vec{x}}
\eqno(1.12)
$$
The equation for the amplitude $a$
$$
\ddot{a} +(\vec{p}^2 +m^2)a =0
\eqno(1.13)
$$
is an equation for linear oscillator with frequency
$$
\omega^2(p) = \vec{p}^2 +m^2 \;\; ,
$$
  or
$$
\omega(p) = \pm\sqrt{p^2 +m^2}
$$

We get that dependence of frequency $\omega$ on $\vec{p}$ and the dependence
of particle energy on momentum $\vec{p}$ are exactly the same (in the units
$\hbar = c =1$). This is why we can use free fields to describe free
particles.

The general solution in the periodic box can be presented as a superposition
of the solutions (1.12)
$$
\Phi(x) = \sum_p [a(p)e^{-ipx} +a^+(p)e^{ipx}]
\eqno(1.15)
$$
where
$$
px = p_{\mu}x_{\mu} = p_0 x_0 -\vec{p}\vec{x}
$$
$$
p_0 = \sqrt{\vec{p}^2 +m^2}
$$
$$
\sum_p = \int\frac{d^3 p}{(2\pi)^3 2p_0}
$$
In the Classical theory coefficient $a(\vec{p})$ are the arbitrary complex
numbers. In terms of these variables the Hamiltonian is equal
$$
H = \int d^3 x {\cal H} = \sum_p \frac{1}{2}\omega(p)[aa^+ + a^+ a]
$$
$$
~~~
\eqno(1.16)
$$
$$
\omega(p) = \sqrt{\vec{p}^2 +m^2}
$$

This is the Hamiltonian for the set of decoupled linear oscillators.

In Quantum Field Theory we have to quantize these oscillators. The variables
$a(p)$ become operators that satisfy commutation relations
$$
[a(\vec{p}), a^+(\vec{p}')] = \delta_{\bar{p}\bar{p}'}
$$
$$
~~~
\eqno(1.17)
$$
$$
[a(\vec{p}), a(\vec{p}')] = [a^+(\bar{p}), a^+(\bar{p}')] = 0
$$
Operators $a(p)$ and $a^+(p)$ are familiar from QM.
They are the annihilation and creation
operator for oscillator with frequency $\omega(\vec{p})$.

The Fock space is the Hilbert space of the states with definite values of the
operator of particle number $N(p) = a^+(p)a(p)$:

\begin{center}

\underline{vacuum $\;\;|0>$}

\end{center}
$$
\left\{
\begin{array}{l}
| 0 > \\
a(p) | 0 > \equiv 0 \;\; ,
\end{array}
\right.
$$

\begin{center}

\underline{one-particle states}

\end{center}
$$
|p> = a^+(p) |0> \;\;
\eqno(1.18)
$$

\begin{center}

\underline{two-particle states}

\end{center}
$$
|p_1, p_2 > = a^+(p_1)a^+(p_2) |0>
$$
$$
|p; p > = \sqrt{2} a^+(p)a^+(p) |0>
$$
etc.

We get that commutation relation eq. (1.17) corresponds to Bose-Einstein
statistic for spin $0$ particle
$$
|p_1, p_2 > = +|p_2, p_1 >
$$
and to positively defined operator of energy
$$
H = \sum \omega(p)[N(p) + \frac{1}{2}]
\eqno(1.20)
$$
The vacuum energy is equal
$$
E_{vac} = \sum \frac{1}{2}\omega(p)
\eqno(1.21)
$$

We have constructed the space of free particles with given momenta. Now we
have to describe the propagation of free particles. The operator
$$
\Phi^{(+)} = \sum_{\bar{p}}a(\bar{p})e^{-ipx}
$$
with positive frequency is a combination of terms that annihilate 1 particle
at point $x$. Operator
$$
\Phi(x) = \sum_{\bar{p}}a^+(\bar{p})e^{ipx}
\eqno(1.22)
$$
creates the particle at point $x$.

Consider the time ordering product
$$
T\{\Phi(x), \Phi(0)\} = \Theta(x_0)\Phi(x)\Phi(0) +\Theta(-x_0)\Phi(0)\Phi(x)
\eqno(1.23)
$$
where the step function is equal
$$
\Theta(x) = \left\{
\begin{array}{lll}
1 & \mbox{\rm if} & x > 0 \\
0 & \mbox{\rm if} & x < 0
\end{array}
\right.
$$

So the Feynman propagator
$$
D_F(x,0) = <0| T\{\Phi(x)\Phi(0)\} |0>
\eqno(1.24)
$$
is the amplitude for a particle to propagate from point $0$ to point $x$.
Time ordering implies that creation always comes before annihilation.

The theory of the \underline{complex scalar fields} $\Phi(x) =
 \frac{1}{\sqrt{2}}(\Phi_1 +i\Phi_2)$ with lagrangian density
 $$
{\cal L} = (\partial_{\mu}\Phi^+\partial_{\mu}\Phi)-m^2\Phi^+\Phi
\eqno(1.25)
$$
is equal to the theory of two different scalar particles with degenerate
masses. The general solution of the field equations can be presented in the
form
$$
\Phi(x) = \sum_p(a(p)e^{-ipx}+b(p)^+ e^{ipx})
\eqno(1.26)
$$
where the operator $(a, a^+)$ and $(b, b^+)$ are creation and annihilation
operators for the particle with the same masses but with the opposite
electric charges (see the next lecture). We consider these two particles as a
particle and antiparticle.

The Feynman propagator of scalar particle in the momentum representation is
equal to

\begin{fmffile}{feyn_graph}

$$
\begin{array}{ll}
\parbox{40pt}{
\begin{fmfgraph*}(40,15) \fmfpen{thin}
\fmfleft{i}\fmfright{o} \fmf{fermion,label=$p$,l.side=left}{i,o}
\end{fmfgraph*}}\;\;\;\;\;\; & 
 D(p) = \frac{i}{p^2 -m^2 +i\varepsilon}\\
\end{array}
\eqno(1.27)
$$\\

 For \underline{Dirac spinor} field $\Psi(x)$ the lagrangian density
 $$
 {\cal L} = \bar{\Psi}[i\gamma_{\mu}\partial_{\mu} -m]\Psi \;\; .
 \eqno(1.28)
 $$

 The plane wave solutions of the Dirac equation look like
$$
u(p, \lambda)e^{ipx} \; , \;\; (\lambda = \pm 1/2)
$$
$$
~~~
\eqno(1.29)
$$
$$
v(p, \lambda)e^{-ipx} \; , \;\; (\lambda = \pm 1/2)
$$
where $u(p, \lambda)$, $v(p, \lambda)$ satisfy equations
 $$
 (\gamma_{\mu}p_{\mu}-m) u(p, \lambda) = 0
 $$
 $$
 ~~~
 \eqno(1.30)
 $$
$$
(\gamma_{\mu}p_{\mu}+m) v(p, \lambda) = 0
$$
and $\lambda = \pm 1/2$ label the independent solution with different value
 of the spin projection on momenta $\vec{p}$. The general solution of Dirac
 equation can be presented in the form
 $$
 \Psi(x) = \sum_{\vec{p}, \lambda}
 \left\{a(p,\lambda)u(p,\lambda)e^{-ipx}+b^+(p, \lambda) v(p, \lambda)e^{ipx}
 \right\}
 \eqno(1.31)
 $$
 where $a(p,\lambda)$ and $b(p,\lambda)$ are annihilation operators for
 particles and antiparticles respectively.

 The next step is the quantization. We have to consider $a(p,\lambda)$ and
 $b(p,\lambda)$ as operators in the Fock space. The great surprise is that to
 have positively defined energy the operators $a(p,\lambda)$ and
 $b(p,\lambda)$ should satisfy \underline{anticommutation} relations
 $$
 \left\{a(p,\lambda), a^+(p',\lambda')\right\} =
 \left\{b(p,\lambda), b^+(p',\lambda')\right\} =
 \delta_{p p'}\delta_{\lambda \lambda'}
 $$
 $$
 ~~~~
 \eqno(1.32)
 $$
 $$
 \{a,a\} = \{a^+, a^+\} = \{b, b\} = \{b^+, b^+\} =0
 $$
with $\{A, B\} = AB +BA$.

 These imply Fermi-Dirac statistic for spin $1/2$ particle. These two
 examples demonstrate the famous spin-statistic theorem.

 Feynman propagator $S_F(x)$
$$
S_F = <0| T\{\Psi(x)\bar{\Psi}(y)\}|0>
\eqno(1.33)
$$
in momentum representation looks like
$$
\begin{array}{ll}
\parbox{40pt}{
\begin{fmfgraph*}(40,15) \fmfpen{thin}
\fmfleft{i}\fmfright{o} \fmf{fermion,label=$p$,l.side=left}{i,o}
\end{fmfgraph*}}\;\;\;\;\;\; & 
S(p) = \frac{i}{\hat{p} -m}
\end{array}
\eqno(1.34)
$$
where $\hat{p} = \gamma_{\mu}p_{\mu}$.

\vspace{3mm}

{\it Exercise.} Calculate the dimension of field $\Psi$: $[\Psi] = m^{3/2}$.

\underline{Electromagnetic Field} $A_{\mu}(x)$:

Lagrangian density is
$$
{\cal L} = -\frac{1}{4} F_{\mu\nu}F_{\mu\nu}+e j_{\mu}^{ext} A_{\mu}
\eqno(1.35)
$$
$$
F_{\mu\nu} = \partial_{\mu}A_{\nu} -\partial_{\nu} A_{\mu}
$$
Because of gauge invariance the quantization of electromagnetic field is
rather subtle matter. In Feynman gauge the propagator for photon
$$
\begin{array}{ll}
\parbox{40pt}{
\begin{fmfgraph*}(40,15) \fmfpen{thin}
\fmfleft{i}\fmfright{o} \fmf{photon,label=$p$,l.side=left}{i,o}
\end{fmfgraph*}}\;\;\;\;\;\; & 
D_F^{\mu\nu} = \frac{(-i)g_{\mu\nu}}{p^2 +i\varepsilon}
\end{array}
\eqno(1.36)
$$

\vspace{3mm}

{\it Exercise.} Show that
$$
[A_{\mu}] = m
$$
$$
[j_{\mu}] = m^3
$$

\vspace{5mm}

{\bf 1.3. Feynman Rules and Feynman Amplitudes.}

\hspace*{9mm} {\bf Tree approximation.}

\vspace{3mm}

What we do understand well is the QFT in the framework of perturbation
theory. Free field theory provides the asymptotic $|{\rm in}>$ and
$|{\rm out}>$ states for free particles and the amplitudes for propagation of
free particles from one space-time point to another point. The nonlinear
interaction term ${\cal L}_{int}$ in perturbation theory provides the
vertices. Combining vertices and propagators one can construct in
perturbation theory the transition amplitude from one asymptotic state to
another one. Let us remind the main steps.

Transitions are described by means of unitary $S$-matrix: $S^+ S =I$
$$
<f|S|i> = <f|i> +(2\pi)^4 i\delta^{(4)}(\Sigma p_f -\Sigma p_i) <f|T|i>
\eqno(1.37)
$$
where $i$ and $f$ refer to initial and final state.

In perturbation theory
$$
S = T exp\{i\int d^4 x{\cal L}_{int}\} =
$$
$$
~~~
\eqno(1.38)
$$
$$
= I+i\int d^4 x{\cal L}_{int}(x) +\frac{i^2}{2}T\{\int d^4 x_1 {\cal L}(x_1);
\int d^4 x_2 {\cal L}(x_2)\} + ...
$$
where the operators of fields are in the \underline{interaction}
representation.

Consider for example the case of QED. The interaction looks like a product of
electromagnetic current and 4-potential

$$
{\cal L}_{int} = j_{\mu}^{em}(x) A_{\mu}(x)
$$
$$
~~~
\eqno(1.39)
$$
$$
j_{\mu}^{em}(x) = (-ie)\{\bar{e}(x)\gamma_{\mu}e(x) - \frac{2}{3}\bar{u}(x)
\gamma_{\mu} u(x) + ... \}
$$
Feynman rules for this QED lagrangian are summarized in Fig. 1.

\vspace{0.75cm}
$$
\begin{array}[10cm]{lcccc}
& initial\;\;\; states & & final\;\;\; states& \\
& & & & \\
electron \;\;\;\;&
\begin{fmfgraph*}(40,15) \fmfpen{thin}
\fmfleft{i}\fmfright{o} \fmf{fermion,label=$p_{-}$,l.side=left}{i,o} \fmfdot{o}
\end{fmfgraph*}
& 
e(p_{-}) \;\;\;\;\;\;\;\;&
\begin{fmfgraph*}(40,15) \fmfpen{thin}
\fmfleft{i}\fmfright{o} \fmf{fermion,label=$p_{-}$,l.side=left}{i,o} \fmfdot{i}
\end{fmfgraph*}
& 
\bar{e}(p_{-})\\
& & & & \\
positron \;\;\;\;&
\begin{fmfgraph*}(40,15) \fmfpen{thin}
\fmfleft{i}\fmfright{o} \fmf{fermion,label=$p_{+}$,l.side=left}{o,i} \fmfdot{o}
\end{fmfgraph*}
& 
\bar{e}(-p_{+}) \;\;\;\;\;\;\;\;&
\begin{fmfgraph*}(40,15) \fmfpen{thin}
\fmfleft{i}\fmfright{o} \fmf{fermion,label=$p_{+}$,l.side=left}{o,i} \fmfdot{i}
\end{fmfgraph*} &
e(-p_{+})\\
& & & & \\
photon \;\;\;\;&
\begin{fmfgraph*}(40,15) \fmfpen{thin}
\fmfleft{i}\fmfright{o} \fmf{photon}{o,i} \fmfdot{o}
\end{fmfgraph*} &
\varepsilon_{\mu}(k) \;\;\;\;\;\;\;\;&
\begin{fmfgraph*}(40,15) \fmfpen{thin}
\fmfleft{i}\fmfright{o} \fmf{photon}{o,i} \fmfdot{i}
\end{fmfgraph*}&
\varepsilon_{\mu}^{\ast}(k)\\
& & & &\\
&
\begin{fmfgraph*}(40,15) \fmfpen{thin}
\fmfleft{i}\fmfright{o} \fmf{photon}{o,i} \fmfdot{i,o}
\end{fmfgraph*}&
\frac{(-ig_{\mu\nu})}{k^2+i\varepsilon} \;\;\;\;\;\;\;\;& &\\
& & & &\\
&
\begin{fmfgraph*}(40,15) \fmfpen{thin}
\fmfleft{i}\fmfright{o} \fmf{fermion}{i,o} \fmfdot{i,o}
\end{fmfgraph*}&
\frac{i}{\hat{p}-m+i\varepsilon} \;\;\;\;\;\;\;\;& &\\
& & & &\\
& 

\begin{fmfgraph*}(40,40) \fmfpen{thin}
\fmfleft{i1,i2}\fmfright{o1} \fmf{fermion}{i2,v1,i1} \fmf{photon}{v1,o1}
\end{fmfgraph*}&
i\gamma_{\mu}Q \;\;\;\;\;\;\;\;& &\\
\end{array}
$$
\begin{center}
Figure 1: Feynman rules of QED.
\end{center}
\vspace{0.5cm}

Using these rules one can easily construct the transition amplitude. Consider as an
example the process $e^+ e^- \to \mu^+ \mu^-$. There is one diagram for this
process \\
\begin{center}
\begin{fmfgraph*}(60,40) \fmfpen{thin}
\fmfleft{i1,i2}\fmfright{o1,o2} \fmf{fermion}{i2,v1,i1}
\fmf{fermion}{o1,v2,o2}
\fmflabel{$e^+$}{i1}
\fmflabel{$e^-$}{i2}
\fmflabel{$\mu^+$}{o1}
\fmflabel{$\mu^-$}{o2}
\fmf{photon}{v1,v2}
\end{fmfgraph*}
\end{center}
\vspace{0.5cm}

The amplitude $T$ is equal
$$
iT(e^+ e^- \to \mu^+ \mu^-) = (-ie)^2 j_{\alpha}^{(e)}
\frac{(-ig_{\alpha\beta})}{q^2} j_{\beta}^{(\mu)}
\eqno(1.40)
$$
where
$$
j_{\alpha}^{(e)} = \bar{e}(-p_+)\gamma_{\alpha}e(p_-)
$$
$$
j_{\beta}^{(\mu)} = \bar{\mu}(k_-)\gamma_{\beta}\mu(-k_+)
$$

This is the example of the amplitude in the lowest order in the coupling
constant $e$. It contains no loops. There is special name for such diagrams
-- tree diagrams.

\vspace{5mm}

{\bf 1.4. Loop corrections.}

\hspace*{9mm} {\bf Propagator corrections in QED.}

\vspace{3mm}

Consider one-loop correction to the photon propagator 

\begin{center}
\begin{fmfgraph*}(60,40) \fmfpen{thin}
\fmfleft{i1}\fmfright{o1} \fmf{photon}{i1,v1}
\fmf{photon}{v2,o1}
\fmf{fermion,left,tension=0.3}{v1,v2,v1}
\end{fmfgraph*}
\end{center}
$$
\delta D_{\mu\nu} = \frac{(-ig_{\mu\alpha})}{q^2}(-i\Pi_{\alpha\beta}(q))
\frac{(-i)g_{\beta\nu}}{q^2}
\eqno(1.41)
$$
where
\vspace{5mm}
$$
\begin{array}[10cm]{lcccc}
(-i)\Pi_{\alpha\beta}(q)\equiv \alpha&
\begin{fmfgraph*}(40,15) \fmfpen{thin}
\fmfleft{i}\fmfright{o} 
\fmf{fermion,left=0.5,label=$p$}{i,o}
\fmf{fermion,left=0.5,label=$p-q$}{o,i}
\fmfdot{i,o}
\end{fmfgraph*}

& \beta =
\end{array}
$$
\vspace{0.5cm}
$$
= e^2 \int \frac{d^4 p}{(2\pi)^4} (-1) Sp \gamma_{\alpha}
\frac{1}{\hat{p}-m+i\varepsilon}
\gamma_{\beta}\frac{1}{\hat{p}-\hat{q} -m+i\varepsilon}
\eqno(1.42)
$$

For large virtual momenta $p$ the loop correction diverges quadratically
$$
\Pi_{\alpha\beta} \simeq e^2 g_{\alpha\beta}\int\limits_0^{\Lambda} \frac{d^4
p}{p^2} \simeq  g_{\alpha\beta}e^2
\Lambda^2 \to \infty
\eqno(1.43)
$$
for $\Lambda\to\infty$.

The one-loop correction to electron propagator\\  

\begin{center}
\begin{fmfgraph}(100,20) \fmfpen{thin}
\fmfleft{i1}\fmfright{o1} \fmf{plain}{i1,v1}
\fmf{plain}{v1,v2}
\fmf{photon,left,tension=0}{v1,v2}
\fmf{plain}{v2,o1}
\end{fmfgraph}
\end{center}

is proportional
to

$$
e^2 \frac{\hat{p}}{p^2 -m^2}\int\frac{d^4 q}{q^4} \sim e^2
\frac{\hat{p}}{p^2 -m^2} \ln \Lambda^2 \to \infty
\eqno(1.44)
$$
for $\Lambda\to\infty$

The vertex

\begin{center}
\begin{fmfgraph*}(60,40) \fmfpen{thin}
\fmfleft{i1,i2}\fmfright{o1} \fmf{plain}{i2,v1,v2,v3,i1} \fmf{photon,tension=0}{v1,v3}\fmf{photon}{v2,o1}

\end{fmfgraph*}
\end{center}

is proportional to

$$
 e^3 \gamma_{\mu} \ln \Lambda^2 \to \infty\\
\eqno(1.45)
$$

for $\Lambda \to \infty$.

These corrections diverge logarithmically.

These are the simplest examples of the problem of divergences in QFT. It was
a great  success of theoretical physics when Dyson, Feynman, Schwinger and
Tomanaga in the late 40th explained how to work with such theories.

\vspace{5mm}

{\bf 1.5. Renormalizable Field Theories QED.}

\vspace{3mm}

The general philosophy of renormalization can be formulated in the following
way:

1) Suppose that we can split quantum fluctuations on the "fast" fluctuations
 (i.e.
with virtual momenta $p>\Lambda$) and on the "slow" ones ($p < \Lambda$), where
$\Lambda$ is \underline{arbitrary} large parameter.

2) Suppose that we can integrate over the "fast" fluctuations even though the
physics at small distances ($p > \Lambda$) can be unknown.

3) For "slow" fluctuations we get "effective field theory" with ${\cal
L}^{eff}(\Lambda)$ or $S^{eff}(\Lambda)$ and with parameters that depend on
cut-off $\Lambda$.

4) Physics of the low-energy processes does not depend on the value of the
$\Lambda$.

5) For special class of renormalizable theories $S^{eff}(\Lambda)$ depends on
\underline{finite} number of parameters and interaction terms.

Consider how this program works in the case of QED. Suppose that effective
lagrangian has the form
$$
{\cal L}(\Lambda) = -\frac{1}{4}(F_{\mu\nu}^B)^2
+\bar{\Psi}_B(i\gamma_{\mu}\partial_{\mu} -m_B)\Psi_B -
$$
$$
-e_B \bar{\Psi}_B \gamma_{\mu}\Psi_B A_{\mu}^B
\eqno(1.46)
$$
where all quantities with label $B$ depend on $\Lambda$. Consider the
scattering of heavy charged particle on the Coulomb center. In this case we
have to sum up all corrections to the photon propagator.

$$
\begin{array}[c]{lllll}

\begin{fmfgraph*}(20,40) \fmfpen{thin}
\fmfleft{i1,i2}\fmfright{o1,o2} 
\fmf{plain}{i2,v1,o2}
\fmf{photon}{v1,v2,v3}
\fmf{phantom}{i1,v3,o1}
\fmfv{decor.shape=cross,decor.size=4thick}{v3}
\end{fmfgraph*}
&
\begin{array}{c}
+\\ [40pt]
\end{array}
&
\begin{fmfgraph*}(20,40) \fmfpen{thin}
\fmfleft{i1,i2}\fmfright{o1,o2} 
\fmf{plain}{i2,v1,o2}
\fmf{photon}{v1,v2}
\fmf{plain,left=0.5,tension=0.2}{v2,v3,v2}
\fmf{photon}{v3,v4}
\fmf{phantom}{i1,v4,o1}
\fmfv{decor.shape=cross,decor.size=4thick}{v4}
\end{fmfgraph*}
&
\begin{array}{c}
+ \\ [40pt]
\end{array}
&
\begin{array}{c}
...\\ [40pt]
\end{array}
\end{array}
$$

As a result at low-energy the amplitude of Coulomb scattering is equal to
$$
T= \frac{e_B^2(\Lambda)}{1-\frac{e_B^2(\Lambda)}{12\pi^2}\ln
\frac{\Lambda^2}{m_e^2}} \cdot \frac{1}{q^2}
\eqno(1.47)
$$
The coefficient in front of $1/q^2$ is by definition the charge of particle
($1/q^2$ corresponds to $1/r$ dependence in the Coulomb law). So we claim
that combination
$$
e_{ph}^2 = \frac{e_B^2(\Lambda)}{1-\frac{e_B^2(\Lambda)}{12\pi^2}\ln
\frac{\Lambda^2}{m_e^2}}
\eqno(1.48)
$$
is the physical charge. It does not depend on $\Lambda$ and in this way we
find $e_B^2(\Lambda)$ as a function of $\Lambda$.

In the similar way one can define the physical electron mass $m_{ph} = m_e$ as
a pole in the exact propagator of the electron.

Now we are able to formulate the main theorem.

{\it{\bf Theorem.}} If we rewrite the amplitudes of all QED processes that depend
on $e_B$, $m_B$ and $\Lambda$ in terms of $e_{ph}$, $m_{ph}$ the dependence
on $\Lambda$ in these amplitudes will disappear for large $\Lambda$!

\vspace{5mm}

{\bf 1.6. Non-renormalizable Theories.}

\vspace{3mm}

   {\bf Fermi Theory} (1934)

\vspace{2mm}

The first theory of weak interactions was formulated by Fermi. It was very
similar to QED. The lagrangian of interaction was equal to a product of two
vector currents. After the discovery of $P$ and $C$ parity violation this
4-fermion theory was modified so that
$$
{\cal L}_W = \frac{G_F}{\sqrt{2}}j_{\alpha}j_{\alpha}
\eqno(1.49)
$$
where $j_{\alpha} = \bar{\nu}_e \gamma_{\alpha}(1+\gamma_5)e + ...$

Remember that
$$
[j] = m^3 \; ;  \;\; [{\cal L}] = m^4
$$
so the Fermi coupling constant has dimension $-2$:
$$
[G_F] = m^{-2}
$$

From the dimensional analysis it is clear that radiative corrections to
4-fermion interaction\\

\begin{center}
\begin{fmfgraph*}(30,40) \fmfpen{thin}
\fmfleft{i1,i2}\fmfright{o1,o2} 
\fmf{plain}{i1,v1,o1}
\fmf{plain,left=1,tension=0.3}{v1,v2,v1}
\fmf{plain}{i2,v2,o2}
\end{fmfgraph*}
\end{center}

should be of the order of

$$
G_F(1+\Sigma(G_F \Lambda^2)^n)(j)^2
$$
where $\Lambda$ is cut-off. It is also clear that 4-fermion interaction can
generate multi-fermion interaction with diverge coupling constant, e.g.
8-fermion interaction\\
\begin{center}

\begin{fmfgraph}(60,60) 
\fmfpen{thin}
\fmfsurroundn{v}{12}
\fmfpolyn{empty,smooth}{e}{4}
\begin{fmffor}{i}{2}{1}{4}
\fmf{plain}{v[i*3-3],e[i],v[i*3-1]}
\end{fmffor}
\fmf{plain}{v[12],e[1],v[2]}
\end{fmfgraph}

\end{center}

$$
\Delta{\cal L}^{eff} = CG_F^4 [\ln \Lambda^2 +\Sigma(G_F \Lambda^2)^n] (j)^4
$$
etc.

In this way we find that ${\cal L}^{eff}$ should depend on infinite number of
terms.

This is the example of non-renormalizable theory. In such theories  we have 
to fix infinite number of terms in ${\cal L}^{eff}(\Lambda)$ at the scale
 $\Lambda$ (i.e. at small distances $x\sim \Lambda^{-1}$) to reconstruct the
amplitudes at low energy. Nobody knows how to work with nonrenormalizable
theories.

\vspace{5mm}

{\bf 1.7. From field theory to cross sections.}

\vspace{3mm}

  The steps from the formal QFT to the numerical predictions for cross 
sections  and for the decay rates are very simple
\begin{center}

Feynman diagrams \\
$\Downarrow$   \\
Amplitude $T$ \\
$\Downarrow$ \\
Probability $\sim \Sigma |T|^2$

\end{center}

For example the Cross Sections are calculated by the formula
$$
d\sigma_{fi} = \frac{1}{2\sqrt{\lambda(s, m_1^2, m_2^2)}} |T_{fi}|^2 d\tau
$$

where
$$
d\tau = (2\pi)^4 \delta^4(p_1 +p_2 -\Sigma p_f)\prod\limits_{j=1}^N
\frac{d^{3} p_j}{(2\pi)^3 2E_j}
$$
is N-particle phase space and
$$
\lambda(s, m_1^2, m_2^2) = 4[(p_1 p_2)^2 - m_1^2 m_2^2 ]
$$

is relativistic flux. 

The Decay Rate is given by formula
$$
d\Gamma = \frac{1}{2E} |T|^2 d\tau
$$

There exists a well developed routine technology of this kind of calculations.
But sometime we can understand the basic properties of the answer without long
 calculations. Consider three examples of the order of magnitude estimates

1. Decay $\mu \to e\tilde{\nu}_e \nu_{\mu}$\\
In this case we know that
$$
[\Gamma] = m \;\; [G_F] = m^{-2} \; ;  \;\;
$$
and that in the limit $m_{\mu} \gg m_e$  nothing should depends on $m_e$.
So on the ground of the dimensional analysis we conclude
$$
\Rightarrow \Gamma(\mu \to e\tilde{\nu}_e \nu_{\mu}) = \frac{1}{192\pi^3} G_F^2
m_{\mu}^5
$$
Certainly the numerical factor is beyond of these order of magnitude estimates
but the dependence on mass is understood well.

2.Cross section  $\nu e \to \nu e$\\
Cross sections have dimension
$$
[\sigma] = m^{-2}
$$
So for very large energy when we can forget about masses

$$
\sigma(\nu_e e \to \nu_e e) = \frac{1}{3\pi} G_F^2 s 
$$
3.Cross section  $e^+ e^- \to \mu^+ \mu^-$

Fine coupling constant is dimensionless. The cross section does not 
depend on masses at high energy. So
$$
\sigma = \frac{4}{3} \frac{\pi\alpha^2}{s} \; , \;\; s \gg m^2
$$
 It is interesting to note that after some training one can restore
the powers of $\pi$ that originate from the phase space. We have
no time for such training.

\newpage

Lecture II. {\bf Symmetries.}

\vspace{5mm}

In the Standard Model the notion of global and local symmetries plays
the very important role.
In this lecture we shall study the different aspects of symmetries in
QFT using a well known physical examples.

\vspace{5mm}

{\bf 2.1. Global symmetries.}

\vspace{3mm}

We start with the simplest $U(1)$ symmetry and consider as an example
the theory of free electrons.
Electrons are described by 4-component complex field
$\psi_i(i=1,2,3,4)$
named bispinor.
The free Lagrangian density has the form

$$
{\cal L}=\bar\psi(x)(i\gamma_\mu\partial_\mu-m)\psi(x)
\eqno(2.1)
$$
where $\bar\psi=\psi^+\gamma_0$ and $\gamma_{\mu}$ are $4\times4$ Dirac
matrices, $m$ is the electron mass.

It is quite evident that $U(1)$ global phase transformations

$$
\begin{array}{lcl}
\psi(x)&\rightarrow&\psi'(x)=e^{i\alpha}\psi(x)\\
\bar\psi(x)&\rightarrow&\bar\psi'(x)=\bar\psi(x)e^{-i\alpha}
\end{array}
\eqno(2.2)
$$
leave Lagrangian (2.1) invariant.
The global symmetry means that the phase of the transformation is the
same for any of space-time points $x$.

A little bit less trivial example is the theory of two complex
self-interacting scalar fields with the degenerate masses

$$
{\cal L}=
        \partial_\mu\Phi^+\partial_\mu\Phi-m^2\Phi^+\Phi-
        \frac{\lambda}{4}(\Phi^+\Phi)^2,
\eqno(2.3)
$$
where $\Phi$ is the two component column (doublet)

$$
\Phi=\left(
\begin{array}{c}
\varphi^+(x)\\
\varphi^0(x)
\end{array}
\right)
\eqno(2.4)
$$
The Lagrangian (2.3) is invariant under global $SU(2)$
rotations of the complex doublet $\Phi$
$$
\begin{array}{lcl}
\Phi(x)&\rightarrow&\Phi'(x)=S\Phi(x)\\
\Phi^+(x)&\rightarrow&(\Phi'(x))^+=\Phi^+(x)S^+
\end{array}
\eqno(2.5)
$$
where $S$ is unitary $2\times2$ matrix
$$
S^+ S=I
$$
$$
{\rm with} \; \; \;  det S = 1
\eqno(2.6)
$$
This matrix $S$ can be represented in the form
$$
S=\exp[i\frac{\tau_a}{2}\alpha^a]
\eqno(2.7)
$$
where $\tau_a=(\tau_1, \tau_2, \tau_3)$ are Pauli matrices

$$
\tau_1=\left(
\begin{array}{ll}
0 & 1\\
1 & 0
\end{array}\right) \ , \
\tau_2=\left(
\begin{array}{ll}
0 & -i\\
i & 0
\end{array}\right) \ , \
\tau_3=\left(
\begin{array}{ll}
1 & 0\\
0 & -1
\end{array}\right)
\eqno(2.8)
$$
that satisfy to $SU(2)$ Lie algebra commutation relations

$$
[\tau_i \ , \ \tau_k]= 2ie_{ik\ell}\tau_\ell
\eqno(2.9)
$$
where $e_{ik\ell}$ is totally antisymmetric tensor with $e_{123}=1$.
Three independent phases $\alpha^a(a=1,2,3)$ do not depend on
space-time.
The Lagrangian density (2.3) describes Higgs scalars in the
Standard Model.

Another useful example is the approximate $SU(2)_L\times SU(2)_R$
symmetry of strong interactions.
This symmetry has been discovered long before the formulation of QCD
in the framework of such general approach as the current algebra.
It provided the first example of nonlinear realization of symmetry in
QFT. It this lecture
we shall formulate this symmetry using the Lagrangian of QCD.

The mass scale of QCD is defined by $\Lambda_{\rm QCD}$ :

$$
\Lambda_{\rm QCD}\simeq 0.5 \ {\rm Gev}
\eqno(2.10)
$$
On the other hand according to particle data group booklet the masses
of up and down quarks are of the order of a few Mev.
So in a good approximation one can take
$$
m_{u,d}=0
\eqno(2.11)
$$
In this limit QCD Lagrangian for $u$ and $d$ quarks can be rewritten as

$$
{\cal L}=\bar\psi_L(i\gamma_\mu{\cal D}_\mu)\psi_L +
        \bar{\psi_R}(i\gamma_\mu{\cal D}_\mu)\psi_R
\eqno(2.12)
$$
where
 $\psi(x)=\left(\begin{array}{c}u(x)\\d(x)\end{array}\right)$
is $SU(2)$ doublet constructed from bispinor $u(x)$ and $d(x)$.
The subscribes $L$ and $R$ means left and right field by definition

$$
\begin{array}{rcl}
u_{L,R}&=&\frac{1}{2}(1\pm\gamma_5)u \nonumber\\
d_{L,R}&=&\frac{1}{2}(1\pm\gamma_5)d
\end{array}
\eqno(2.13)
$$
where $\gamma_5= -i\gamma_0\gamma_1\gamma_2\gamma_3$.

Any bispinor $\psi(x)$ can be represented as a sum of two spinors
(left and right Weyls spinors)

$$
\psi=\psi_L+\psi_R=\frac{1}{2}(1+\gamma_5)\psi+\frac{1}{2}(1-\gamma_5)\psi
\eqno(2.14)
$$
For the case of massless particles the Lagrangian itself can be written
as a sum of two independent Lagrangians : one for left spinors,
another for right spinors.
Each of them is $SU(2)$ invariant.
So the total symmetry is $SU(2)_L \otimes SU(2)_R$.

\vspace{5mm}

{\bf 2.2. Noether's Theorem. Conserved Currents.}

\vspace{3mm}

So far we considered the examples of different Lagrangians that were
invariant with respect to $U(1)$ and $SU(2)$ transformations.
The transformations were global, i.e. they did not depend on
space-time points $x$.
As for invariance of the Lagrangian it looked like rather trivial
property.

Noether's theorem states that for any continuous global symmetry of
the Lagrangian one can construct the conserved vector currents.
This is dynamical statement and it does not look like being trivial at
all.
I am going to prove this theorem in the classical field theory.

Let Lagrangian $L$ depends on the set of fields $\varphi^i$ and
its first derivatives $\varphi^i_{,\mu}=\partial_\mu\varphi^i$.
For infinitesimal global transformations the variations of fields are
equal to

$$
\begin{array}{rcl}
\delta\varphi^i&=&i\epsilon^{(a)}T^{(a)}_{ij}\varphi^j \nonumber\\
\delta\varphi^i_{,\mu}&=&i\epsilon^{(a)}\partial_\mu(T^a_{ij}\varphi^j)
\end{array}
\eqno(2.15)
$$
The real infinitesimal parameters $\epsilon^{(a)}$ represent the
independent symmetry transformations, matrices $T^a_{ij}$ are the
generators of the group of transformations in given representations.
The invariance means that the actions $S$ is not changed under
transformation (2.15):

$$
\delta S=\int d^4x\delta L \equiv 0
\eqno(2.16)
$$
Let us calculate the variation of Lagrangian density directly

$$
\begin{array}{rcl}
\delta L&=&\frac{\partial L}{\partial\varphi^i}\delta\varphi^i+
        \frac{\partial L}{\partial \varphi^i_{,\mu}}
        \delta\varphi^i_{,\mu}= \nonumber\\
&=&+[\partial_\mu\frac{\partial L}{\partial\varphi^i_{,\mu}}]
        \delta\varphi^i+\frac{\partial L}{\partial\varphi^i_{,\mu}}
        \delta\varphi^i_{,\mu}
\end{array}
\eqno(2.17)
$$
where we have used the Lagrangian equation of motion for
$\varphi^i(x)$

$$
\frac{\partial L}{\partial \varphi^i}=
\partial_\mu\frac{\partial L}{\partial\varphi^i_{,\mu}}
\eqno(2.18)
$$
Substituting the variations for $\delta\varphi$ and
$\delta\varphi_{,\mu}$ from eqs. (2.15) into (2.17) and
(2.16) we get

$$
\delta S=i\epsilon^{(a)}\int d^4 x \partial_\mu j^a_\mu (x) =0 \;\; ,
\eqno(2.19)
$$
where
$$
j^{(a)}_\mu =
\frac{\partial L}{\partial\varphi^i_{,\mu}}T^{(a)}_{ij}\varphi^j
\eqno(2.20)
$$
Therefore we get the conservation of Noether currents

$$
\partial_\mu j^{(a)}_\mu (x)=0
\eqno(2.21)
$$
and the conservation of the corresponding charges
$$
\frac{d}{dt}Q^{(a)}(t)=0
\eqno(2.22)
$$
$$
Q^{(a)}(t)=\int d^3xj^{(a)}_0(x)
\eqno(2.23)
$$
(We assume that there are no fields at spatial infinity,
i.e. $\int d^3x\partial_ij_i\equiv 0$)

The generalization of this proof to the Quantum Field Theory requires more
advanced techniques such as operators algebra, commutators etc.
But the final results, i.e. the expression for conserved Noethers
currents, remain the same.
So we are going to skip the proof of the theorem in QFT.

At the end of this subsection we present the conserved vector currents
that correspond to the symmetries that we considered in the subsection
2.1 :

$$
\begin{array}{rcl}
U(1)&:&j_\mu=\bar\psi\gamma_\mu\psi \ \ \ , \nonumber\\
SU(2)&:&j_\mu^a=\Phi^+\tau^a\stackrel{\leftrightarrow}{\partial_\mu}\Phi
\end{array}
\eqno(2.24)
$$

$SU(2)_L\times SU(2)_R$ :
$$
(j_\mu^{L,R})^a=\bar\psi_{L,R}\gamma_\mu\tau^a\psi_{L,R}
$$
where
$$
\Phi^+\stackrel{\leftrightarrow}{\partial_\mu}\Phi=
        \Phi^+\partial_\mu\Phi-(\partial_\mu\Phi^+)\phi
$$

\vspace{5mm}

{\bf 2.3. Spontaneous Violation of Global Symmetry.}

\hspace*{9mm} {\bf  Goldstone Phenomenon.}

\vspace{3mm}

The idea of spontaneous violation of symmetry was formulated first in
the solid state physics.
Consider, for example, the piece of some ferromagnetic material.
The interaction of the elementary magnetic moments of electrons inside
ferromagnetic is $O(3)$ invariant.
On the other hand at low temperature $T<T_c$ the total magnetic moment
$\overrightarrow M$ of ferromagnetic piece is nonzero and has the
definite direction, i.e. it violates $O(3)$ invariance of the system.
Ground state is only $O(2)$ invariant for the rotations around
$\overrightarrow M$ and the ``violated'' symmetries are realized as a
massless excitations.

In field theory analogous phenomenon is known as Nambu-Goldstone
realization of symmetry.
We are going to discuss this phenomenon using the example of $SU(2)_L
\times SU(2)_R$ symmetry of strong interactions.
Quark Lagrangian (2.12) is $SU(2)_L\times SU(2)_R$ invariant.
There are 3 left $(V-A)$ and 3 right $(V+A)$ conserved currents of
massless quarks.
In other words there are 3 vector and 3 axial vector
conserved currents.
This is evident.
On the other hand the quarks do not exist like a free particles.
Instead we have a set of massive hadrons -- baryons and mesons.
The $SU(2)_V$ symmetry of hadrons was known for a long time.
For example proton and neutron have practically degenerate mass and can
be treated as an up and down members of $SU(2)_V$ doublet
$N=(\stackrel{p}{_{n}})$.
(There are small corrections to the $SU(2)$ symmetric approximation
of the order of $(\frac{m_{u,d}}{\Lambda})^2$ and of the order of
electroweak coupling constant $\alpha$.)
The matrix elements of the conserved $SU(2)$ vector currents between
nucleon states have the form

$$
<N|\bar\psi\gamma_\mu\tau^a\psi|N>\sim\bar N\gamma_\mu\tau^aN
\eqno(2.25)
$$
In the limit of degenerate mass this m.e. is transversal 
$$
q_\mu \bar N\gamma_\mu \tau^aN=0 \ \ \ ,
\eqno(2.26)
$$
i.e. it corresponds to conserved currents.

For conservation it is crucial to have baryons with
\underline{degenerate}
masses.
The symmetry is realized in such a way that it transforms one-particle
baryonic state into another one-particle state with the same
mass.

If this realization of symmetry is unique we immediately get troubles
with the $SU(2)_A$ symmetry.
Indeed to construct the transversal matrix element of axial current we need
degenerate baryons with opposite $P$-parity.
The brief search for such baryons in the Table of Particle Properties
shows that such baryons do not exist in Nature.
So this way is prohibited for $SU(2)_A$.

One may try another possibility for matrix element of axial current
between the same nucleon states
$$
<N|\bar\psi\gamma_\mu\gamma_5\tau^a|N>\sim
        (g_{\mu\nu}-\frac{q_\mu q_\nu}{q})
        \bar N\gamma_\nu\gamma_5\tau^aN \ ,
\eqno(2.27)
$$
where $q$ is momentum transfer from one nucleon to another.
The transversality is now evident because
$$
q_\mu(g_{\mu\nu}-\frac{q_\mu q_\nu}{q^2})\equiv 0
$$

So we have solved this problem easily.
But now matrix element (2.27) is singular.
It has a pole at $q^2=0$.
The poles in the amplitudes correspond to one particle intermediate
states and pole at $q^2=0$ corresponds to massless particle.
Fortunately there are $\pi$-mesons that contribute into m.e. (2.27) and
that are almost massless.
Now we are able to formulate the new way of realization of approximate
$SU(2)_A$.

According to this new philosophy $\pi$-mesons should be massless in
the limit $m_{u,d}=0$ and there should be simple relation between the
axial-vector part of matrix element
$$
g_{\mu\nu}\bar N\gamma_\mu\gamma_5 N
\eqno(2.28)
$$
and the pseudoscalar part

$$
-\frac{q_\mu q_\nu}{q^2}\bar N\gamma_\mu\gamma_5 N=
-\frac{q_\mu}{q^2}2m_N(\bar N\gamma_5N)
\eqno(2.29)
$$
to have conserved current.
This relation is known as Goldberger-Treiman relation.
The matrix element of axial current between nucleon states can be
measured in $n\rightarrow pe\tilde\nu$ decay.
Experimental value of the ratio of axial coupling constant to the
pseudoscalar coupling constant is very close to the theoretical
prediction.
So this realization of symmetry indeed works in
the case of $SU(2)_A$.

Instead of one-particle degenerate states with opposite
$P$-parity we have massless pseudoscalar Goldstone particles - pions.
The symmetry transforms one-particle baryonic state into degenerate
two-particle state (baryon plus massless pseudoscalar pion), then
into 3-particle state with two pions etc.
This is nonlinear realization of symmetry.
One can proceed further and show that there are infinitely many states
with the same lowest energy. The vacuum state
is one of these states and it violates $SU(2)_A$
symmetry.
This violation is due to nonzero vacuum expectation value of quark
condensate

$$
\begin{array}{rcl}
<0|\bar u u | 0>&=&<0|\bar d d|0>\neq 0 \nonumber\\
<0|\bar u\gamma_5 u|0>&=&<0|\bar d \gamma_5d|0>=0
\end{array}
\eqno(2.30)
$$
It seems more instructive not to spend time proving eqs. (2.30)
but to consider the same phenomenon using a very simple field model
studied many years ago by Goldstone.

Let us start with the theory of complex scalar field $\varphi(x)$ with
Lagrangian

$$
{\cal L}=\partial_\mu\varphi^+\partial_\mu\varphi-V(|\varphi|^2)
\eqno(2.31)
$$
and with a special choice of potential (see fig.1)
$$
V(|\varphi|^2)=\lambda(|\varphi|^2-\frac{\eta^2}{2})^2
\eqno(2.32)
$$

\begin{figure}
\begin{center}
\epsfbox{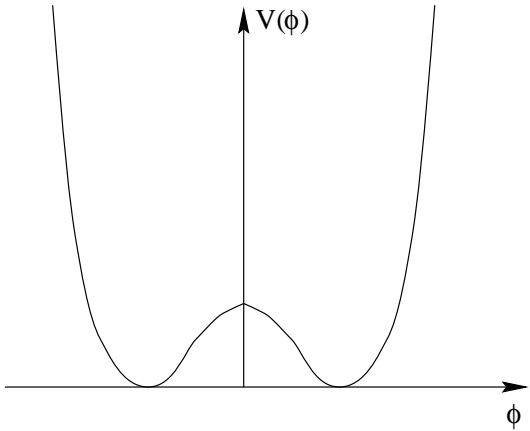}
\caption{ Higgs Potential in Standard Model}
\end{center}
\end{figure}

Lagrangian (2.31) is invariant under $U(1)$ transformations

$$
\varphi(x)\rightarrow \varphi'(x)=e^{i\Lambda}\varphi(x)
\eqno(2.33)
$$
and the Noether current is

$$
j_\mu=i\varphi^+\stackrel{\leftrightarrow}{\partial_\mu}\varphi
\eqno(2.34)
$$

There are continuously many minima of the potential $V$ (2.32)

$$
\varphi=\frac{1}{\sqrt{2}}\eta e^{i\alpha}
\eqno(2.35)
$$
The vacuum  corresponds to one of these minima.
This is spontaneous violation of symmetry : we have chosen one of
state as a vacuum from the infinite set of minima.
Let vacuum state corresponds to zero phase $\alpha=0$ :

$$
\varphi_{cl}=\frac{1}{\sqrt{2}}\eta
\eqno(2.36)
$$

Consider the small fluctuation of fields near vacuum configuration

$$
\varphi=\frac{1}{\sqrt{2}}[\eta+\rho(x)+i\sigma(x)]
\eqno(2.37)
$$

Potential can be rewritten as

$$
V(\varphi)=V(\rho,\sigma)=\frac{\lambda}{2}\left\{(\sigma^2+\rho^2)^2+
        4\eta \rho(\rho^2+\sigma^2)+
        4\eta^2\rho^2\right\}
\eqno(2.38)
$$

The coefficients in front of bilinear terms determine the mass of the fields.
So we get a theory of two particles with masses
$$
\begin{array}{c}
M^2_\rho=4\lambda\eta^2\nonumber\\
M^2_\sigma\equiv 0
\end{array}
\eqno(2.39)
$$

Excitations that correspond to the motion along the valley of minima
are massless!
This is Goldstone phenomenon.

We can use more elegant and transparent representation for
$\varphi(x)$ to demonstrate this phenomenon.
Let us rewrite $\varphi(x)$ in terms of modulus and phase

$$
\varphi(x)=\rho(x)e^{i\sigma(x)}
\eqno(2.40)
$$
Then

$$
{\cal L}(\rho,\sigma)=
(\partial_\mu\rho)^2-V(\rho^2)+\rho^2(\partial_\mu\sigma)^2
\eqno(2.41)
$$

There is no dependence on the field $\sigma$ in the potential
and therefore this field corresponds to massless particle.

In Quantum Field Theory we have two ways for realization of
symmetry:

\begin{itemize}
\item[1)]Vacuum state has the symmetry of the action $S$.
        Excitation states are degenerate.
\item[2)]Vacuum state has lower symmetry than action $S$.
        There are flat direction in configuration space of fields.
        The motions along these flat directions correspond to
        massless Goldstone particles.
\end{itemize}

\vspace{3mm}

{\it Exercise}

Consider the double well potential in Quantum Mechanics
$$
V(x)=\lambda(x^2-\eta^2)^2
$$
and in the Quantum Field Theory for real scalar field $\varphi(x)$ ;

$$
V(\varphi)=\lambda(\varphi^2-\eta^2) \ .
$$
Show that in QM there is only one lowers state and in QFT there are two
degenerate and orthogonal lowest states.

\vspace{5mm}

{\bf 2.4. Local U(1) gauge symmetry.}

\vspace{3mm}

Now we are going to study the new type of symmetries : local gauge
symmetries.
Let us start with the theory of complex field $\varphi(x)$ described
by the Lagrangian (eq.(2.31))

$$
{\cal L}=\partial_\mu\phi^+\partial_\mu\phi-V(\phi^+\phi)
$$
invariant under global $U(1)$ transformation

$$
\phi(x)\rightarrow\phi'(x)=e^{i\Lambda}\phi(x) \ .
$$
Consider now the local $U(1)$ transformation when we change the phase
of the field independently for any point $x$

$$
\phi(x)\rightarrow\phi'(x)=e^{i\Lambda(x)}\phi(x)
\eqno(2.42)
$$
The potential $V(|\phi|^2)=V(|\phi'|^2)$ is invariant under this
transformation but the kinetic term is not

$$
\partial_\mu\phi^+\partial_\mu\phi\rightarrow
\vert(\partial_\mu+i(\partial_\mu\Lambda)\phi\vert^2
\eqno(2.43)
$$
To \underline{compensate} this non-invariant change one can introduce
new \underline{vector} field $A_\mu(x)$ with the transformation law :

$$
A_\mu(x)\rightarrow A'_\mu(x)=A_\mu(x)+\frac{1}{e}\partial_\mu\Lambda(x)
\eqno(2.44)
$$

so that the new Lagrangian

$$
{\cal L}=\vert(\partial_\mu-ieA_\mu)\phi\vert^2 -V(|\phi|^2)
\eqno(2.45)
$$
is locally $U(1)$ invariant or gauge invariant.
The combination ${\cal D}_\mu=\partial_\mu-ieA_\mu$ has a name of
covariant derivative (or long derivative).
It has a simple transformations law

$$
\begin{array}{rcl}
D_\mu&\rightarrow&e^{i\Lambda}{\cal D}_\mu e^{-i\Lambda}\nonumber\\
D_\mu\phi&\rightarrow& e^{i\Lambda}({\cal D}_\mu \phi)
\end{array}
\eqno(2.46)
$$

Up to now the fields $A_{\mu}(x)$ have no kinetic term in the Lagrangian
and they are some kind of the auxiliary fields that do not propagate.
To construct kinetic term we need gauge invariant combination of the
derivatives of field $A_{\mu}$. Notice that covariant derivatives and
any combinations of
the covariant derivatives have a very simple transformation law eq. (2.46).
Consider the
commutator of two derivatives,

$$
\begin{array}{l}
[{\cal D}_\mu{\cal D}_\nu]\equiv -ieF_{\mu\nu}\nonumber\\
F_{\mu\nu}=\partial_\mu A_\nu -\partial_\nu A_\mu
\end{array}
\eqno(2.47)
$$
We see that commutator is not the the differential operator but the function 
of $x$. According to (2.46) it is gauge invariant function.
Now we are in position to write the totally gauge invariant Lagrangian

$$
\begin{array}{rcl}
{\cal L}&=&-\frac{1}{4}F_{\mu\nu}F_{\mu\nu}+|{\cal D}_\mu\phi|^2-
        V(|\phi|^2)\nonumber\\ \\
\phi(x)&\rightarrow&\phi'(x)=e^{i\Lambda(x)}\phi(x)\\ \\
A_\mu(x)&\rightarrow&A'_\mu(x)=A_\mu(x)+
\frac{1}{e}\partial_\mu\Lambda(x)\nonumber
\end{array}
\eqno(2.48)
$$
The notion of gauge invariance was introduced by V. Fock in 1926 and
by H. Weyl in 1929.
(The very interesting history of this subject can be found in the lectures 
given by L.Okun at this school ten years ago).

\vspace{5mm}

{\bf 2.5. Spontaneous Violation of local symmetry.}

\hspace*{9mm} {\bf Higgs Phenomenon.}

\vspace{3mm}

For the case when time derivative is zero $D_o\phi=0$ and electric field
is zero $F_{oi}=E_i=0$ the Lagrangian (2.48) formally is equal to the free
energy in the Ginzburg-Landau
phenomenological theory of superconductivity,
where $\varphi(x)$ plays a role of the order parameter.
It is known that  magnetic field does not
penetrate into superconductor, it falls exponentially.
Exponential fall in QFT corresponds to a massive particle.
So one can expect that Lagrangian (2.48) at certain circumstances can describe 
the massive gauge field.
This is the famous Higgs mechanism of spontaneous violation of local
symmetry.

Consider Lagrangian (2.48) with the special choice of
potential energy (2.32)

$$
{\cal L}=-\frac{1}{4}F^2_{\mu\nu}+|{\cal D}_\mu\phi|^2-
        \lambda(|\phi|^2-\frac{\eta^2}{2})^2
\eqno(2.49)
$$

Potential $V(\phi)$ has continuous valley of minima.
Let us quantize the fields near the vacuum state (2.32)

$$
<\varphi>=\varphi_{cl}=\frac{1}{\sqrt{2}}\eta \nonumber
\eqno(2.50)
$$

As in the case of global symmetry it is convenient to use representation
of $\phi(x)$ in term of modulus and phase

$$
\phi(x)=\frac{1}{\sqrt{2}}(\eta+\rho(x))e^{i\sigma(x)}
\eqno(2.51)
$$
The Lagrangian (2.48)-(2.49) is gauge invariant.
So let us make gauge transformation with $\Lambda(x)\equiv-\sigma(x)$

$$
\begin{array}{rcl}
\phi(x)&\rightarrow&\phi'=e^{i\sigma}\phi\nonumber\\
A_\mu(x)&\rightarrow&A'_\mu=A_\mu-\frac{1}{e}\partial_{\mu}\sigma
\end{array}
\eqno(2.52)
$$
In this gauge (unitary gauge)

$$
D_\mu\phi\rightarrow
(\partial_\mu-ieA_\mu)\frac{1}{\sqrt{2}}(\eta+\rho(x))
\eqno(2.53)
$$
and the Lagrangian can be rewritten in the form

$$
\begin{array}{rcl}
{\cal L}&=&
        \left[-\frac{1}{4}F^2_{\mu\nu}+\frac{1}{2}
                e^2\eta^2 A^2_{\mu}\right]+\nonumber\\
&&\frac{1}{2}(\partial_\mu\rho)^2+(e^2\eta)\rho(x)A^2_\mu(x)+
        \frac{e^2}{2}A^2_\mu(x)\rho^2(x)
\end{array}
\eqno(2.54)
$$
The term in bracket represents the free massive vector particle with
mass

$$
m_V=e\eta
\eqno(2.55)
$$
Massless Goldstone mode $\sigma(x)$ has been eaten by massless vector
field $A_\mu(x)$ (that had two polarization) and as a result we get
massive vector field with three polarization.
This is Higgs Phenomenon.

\vspace{5mm}

{\bf 2.6. Local $SU(2)$. Yang-Mills theory of vector fields.}

\vspace{3mm}

We have to make another nontrivial step to be ready for the
construction of the Standard Model.
We have to consider the general case of the local gauge groups.

Let us start with $SU(2)$ theory of massless fermion
$\psi=\left(
\begin{array}{c}
\psi_1\\
\psi_2
\end{array}\right)$

$$
{\cal L}=\bar\psi[i\gamma_\mu\partial_\mu\psi]
\eqno(2.56)
$$
and consider \underline{local} $SU(2)$ transformations

$$
\psi(x)\rightarrow\psi'(x)=S(x)\psi(x)
\eqno(2.57)
$$
where
$$
\begin{array}{ll}
&S(x)=\exp i(T_j\Lambda_j(x)) \ ; \nonumber\\
&T_i=\frac{1}{2}\tau_i \ \ , \ \ i = 1,2,3 \ ; \\
&[ T_i \ , \  T_j]=ie_{ijk}T_k \nonumber
\end{array}
\eqno(2.58)
$$
The Lagrangian (2.56) is \underline{not} invariant under this
transformation.
To compensate the non-invariant piece in the Lagrangian we introduce the
\underline{triplet} of vector fields $A^i_\mu(x)$ so that:

$$
\begin{array}{l}
{\cal L}=\bar\psi i\gamma_\mu(\partial_\mu - igA_\mu(x))\psi\nonumber\\
A_\mu(x)= T^iA^i(x)
\end{array}
\eqno(2.59)
$$
with the transformation law

$$
A_\mu(x)\rightarrow A'_\mu(x)=SA_\mu(x)S^+-\frac{i}{g}(\partial_\mu S)S^+
\eqno(2.60)
$$
Again it is convenient to introduce covariant derivative

$$
{\cal D}_\mu=\partial_\mu-igA_\mu
\eqno(2.61)
$$
that transforms as a triplet under $SU(2)$ transformations:
$$
\begin{array}{ll}
D_\mu\rightarrow&SD_\mu S^+\nonumber\\
D_\mu\psi\rightarrow&S(D_\mu\psi)
\end{array}
\eqno(2.62)
$$
We can define the triplet of
field-strength tensor $G^i_{\mu\nu}$ :

$$
\begin{array}{rl}
G_{\mu\nu}&\equiv G^i_{\mu\nu}T^i=\frac{i}{g}
        [{\cal D}_\mu , {\cal D}_\nu]\nonumber\\
&=\partial_\mu A_\nu-\partial_\nu A_\mu -ig[A_\mu A_\nu]\\
&G_{\mu\nu}\rightarrow G'_{\mu\nu}=SG_{\mu\nu}S^+
\end{array}
\eqno(2.63)
$$
and construct the $SU(2)$ gauge invariant Lagrangian

$$
{\cal L}=-\frac{1}{4}Tr[G_{\mu\nu}G_{\mu\nu}]+
        \bar\psi i\gamma_\mu{\cal D}_\mu \psi
\eqno(2.64)
$$
This Lagrangian was invented by Yang and Mills in 1954.
The very nontrivial part in this construction is that kinetic energy
$\sim G^2_{\mu\nu}$ contains bilinear $\sim A^2$, trilinear $\sim A^3$
and quadralinear $\sim A^4$ terms:

$$
\begin{array}{ll}
\begin{fmfgraph*}(40,40) \fmfpen{thin}
\fmfleft{i1,i2}\fmfright{o1,o2} 
\fmf{phantom}{i2,v1,o2}
\fmf{photon}{v1,v2}
\fmf{photon}{i1,v2,o1}
\fmflabel{$A$}{v1}
\fmflabel{$A$}{i1}
\fmflabel{$A$}{o1}
\end{fmfgraph*}\;\;\;\;\;\; & 
\begin{fmfgraph*}(40,40) \fmfpen{thin}
\fmfleft{i1,i2}\fmfright{o1,o2} 
\fmf{photon}{i1,o2}
\fmf{photon}{i2,o1}
\fmflabel{$A$}{i1}
\fmflabel{$A$}{i2}
\fmflabel{$A$}{o1}
\fmflabel{$A$}{o2}
\end{fmfgraph*}
\end{array}
$$
\vspace{8mm}

So we have a gauge theory of self-interacting vector fields.

\vspace{5mm}

{\bf 2.7. Spontaneous Violation of Local {\boldmath $SU(2)$}
Symmetry.}

\hspace*{9mm}{\bf Renormalizable theory of massive vector fields.}

\vspace{3mm}

Consider the $SU(2)$ gauge theory of the couple of scalar fields
$\phi=\left(
\begin{array}{c}
\varphi_1\\
\varphi_2
\end{array}\right)$:

$$
{\cal L}=-\frac{1}{4}TrG_{\mu\nu}G_{\mu\nu}+|{\cal D}_\mu\phi|^2-
        \lambda(|\varphi|^2-\frac{\eta^2}{2})^2
\eqno(2.65)
$$
We expect that after spontaneous violation of $SU(2)$
symmetry three Goldstone
bosons will be mixed with three massless vector fields and produce
three massive vector filed.
To show this we  repeat the steps that we had done in the case
of local $U(1)$ symmetry.

Let us introduce a special representation for the doublet $\phi$

$$
\phi(x)=\left(
\begin{array}{c}
\varphi_1\\
\varphi_2
\end{array}\right)= e^{i\sigma^i(x)T^i}\left(
\begin{array}{l}
0\\
\frac{1}{\sqrt{2}}(\eta+\rho(x))
\end{array}\right)
\eqno(2.66)
$$
and consider gauge transformation with the parameter
$$
\Lambda^i(x)=-\sigma^i(x)
\eqno(2.67)
$$
In this gauge the fields $\sigma^i(x)$ disappear from the Lagrangian and
vector part of $\cal L$ gets the form

$$
\left\{
\begin{array}{l}
{\cal L}_{\rm vect}=-\frac{1}{4}Tr G^2_{\mu\nu}-
        \frac{1}{2}m^2_V A^2_\mu\nonumber\\
m_v=\frac{1}{2}g\eta
\end{array}\right.
\eqno(2.68)
$$
This is the theory of massive vector fields with the special choice of
self-interactions.

The theory of massless Yang-Mills field was renormalizable theory.
It seems that the property of the vacuum should not change the
behavior of the amplitudes at high energy.
So one can believe that Yang-Mills theory with spontaneous violation of
gauge symmetry remains renormalizable.
The theory of massive vector fields with arbitrary interactions is
nonrenormalizable in general.
But if one takes the special case of interaction with quarks, with
scalars and self-interaction that corresponds to the gauge-invariant
Lagrangian (2.65) the nonrenormalizable divergences should
disappear.
Technically the rigorous proof of this statement is quite nontrivial
business even now.
This problem had been solved by t'Hooft and Veltman in 1971.

\newpage

Lecture III. {\bf SU(2)$\times$U(1) Theory of Electroweak Interactions.}

\vspace{5mm}

In this lecture I am going to describe the fundamental Lagrangian of the
Standard Model. It is important to understand that {\bf a priori} there is no
unique way to construct the model of electroweak interactions. There are
plenty of them. In the review paper by B.Bjorken and Llewellyn-Smith in 1973
they discussed several dozens of models. We do not understand yet why the
gauge group is SU(2)$\times$U(1), why there are three generations of quarks
and leptons etc. We have to deduce our theory from the experiment.

\vspace{5mm}

{\bf 3.1. Minimal group.}

\vspace{3mm}

It was well established in old four-fermionic theory of weak interactions
that charged currents (responsible for  $\beta$-decay of nucleons and
other hadrons) have $V-A$ structure, i.e. they are constructed from the
left-handed fermions.

The minimal group of gauge symmetry which includes charged vector currents is
$SU(2)$ group. So any theory of weak interactions have to include $SU(2)_L$
symmetry as a subgroup. Photon interacts both with left- and right-handed
fermions. So if we are going to unify weak and electromagnetic interactions
the group of gauge symmetry should include $U(1)$ as well. The simplest
choice of such group of symmetry is
$$ G = SU(2)_L \times U(1)
$$

\vspace{5mm}

{\bf 3.2. Left and Right Fermions.}

\hspace*{9mm}{\bf {\boldmath$SU(2)_L$} symmetry. Weak Mixing Angles.}

\vspace{3mm}

Any Dirac 4-spinor $\Psi$ can be presented as a sum of two Weyl spinors
$\Psi_L$ and $\Psi_R$:
$$
\Psi = \Psi_L \oplus\Psi_R = \frac{1}{2}(1+\gamma_5)\Psi
+\frac{1}{2}(1-\gamma_5)\Psi \eqno(3.1)
$$

These Weyl spinors are irreducible representation of Lorentz group, they
depend on two complex parameters. For massless fermions
$$
\Psi_L = \left(
\begin{array}{c}
(1-\vec{\sigma}\vec{n})\varphi \\
-(1-\vec{\sigma}\vec{n})\varphi
\end{array}
\right) \;\; , \eqno(3.2.)
$$
where $\varphi$ is 2-spinor, $\vec{\sigma}$ are Pauli matrices, and $\vec{n} =
\vec{p}/|p|$ is the direction of the motion of particle. So for left particle
$\Psi_L$
$$
\vec{\sigma}\vec{n} = -1 \eqno(3.3)
$$
and for right particles $\Psi_R$
$$
\vec{\sigma}\vec{n} = +1 \eqno(3.4)
$$

Left leptons and quarks group into $SU(2)_L$ doublets. For the first
generations they are
$$
L = \left(
\begin{array}{c}
\nu_e \\
e
\end{array}
\right)_L \;\;\; \mbox{\rm and} \;\;\; Q = \left(
\begin{array}{c}
u \\
d
\end{array}
\right)_L  \eqno(3.5)
$$

To avoid $V+A$ charged current we have to put right fermions into singlet
representation. So $e_R$, $u_R$ and $d_R$ are singlets.
As for right-handed neutrino $\nu_R$
nobody has observed it so far. It is unknown whether such field exists.
Just now we prefer not to introduce $\nu_R$ into the theory.

To include the electromagnetic interactions we have to define charge. For
left-handed fermions
the charge is different for up and down component so that
$$
Q_L = T_3 +Y_L \eqno(3.6)
$$
where $T_3$ is the third component of $SU(2)_L$ and $Y_L$ is left
hypercharge.

From eq.(3.6) it follows that for leptonic doublet $Y_L = -1/2$ and for quark
doublet $Y_Q = 1/6$.

For right fermions we identify $Q$ and $Y_R$:
$$
Q_R = Y_R \eqno(3.7)
$$
so that $Y_{u_R} = \frac{2}{3}$; $Y_{d_R} = -\frac{1}{3}$, $Y_{e_R} = -1$.

The minimal way to introduce $U(1)$ interactions is to consider gauge boson
that  interacts with
$$
Y = Y_L +Y_R \eqno(3.8)
$$
This is the gauge group of Minimal Standard Model
$$
SU(2)_L \times U(1)_Y \;\; . \eqno(3.9)
$$
Let $A_{\mu}^i(x)$, $i=1,2,3$ be gauge bosons of $SU(2)_L$ and $B_{\mu}(x)$
-- the gauge boson of $U(1)$ group. The charged fields
$$
A_{\mu}^{\pm} = \frac{1}{\sqrt{2}}(A_{\mu}^1 \pm iA_{\mu}^2) \eqno(3.10)
$$
can be identify with $W_{\mu}^{\pm}$ bosons.

Photon $A_{\mu}(x)$ is in general some combination of $A_{\mu}^3$ and
$B_{\mu}$. Orthogonal combination represents another physical particle that
we identify with $Z$ boson. So
$$
\left(
\begin{array}{c}
Z_{\mu} \\
A_{\mu}
\end{array} \right) = \left(
\begin{array}{cr}
\cos\theta_W & -\sin\theta_W \\
\sin\theta_W & cos\theta_W
\end{array} \right) \left(
\begin{array}{c}
A_{\mu}^3 \\
B_{\mu}
\end{array} \right)  \eqno(3.11)
$$
where $\theta_W$ is a weak mixing angle.

To violate spontaneously $SU(2)_L \times U(1)_Y$ group and to make masses to
$W^{\pm}$ and $Z$ bosons we need three Goldstone fields. The $SU(2)$ doublet
of Higgs particles
$$
H = \left(
\begin{array}{c}
H^+ \\
H^0
\end{array} \right) \; ; \;\;\; Y_H = \frac{1}{2}  \eqno(3.12)
$$
can provide this number of Goldstone bosons after spontaneous violation. In
the MSM we use only {\bf one} Higgs doublet.

We have completed the construction of the MSM. Now we are ready to calculate
the masses of vector bosons $m_W$, $m_Z$ and phenomenological mixing angle
$\theta_W$ in terms of coupling constants $g_2$ for $SU(2)$, 
$g_1$ for $U(1)$  and in terms of v.e.v. of Higgs field $\eta$.

In the unitary gauge Higgs doublet has the form
$$
H(x) = e^{i\vec{T}\vec{\alpha}(x)} \left(
\begin{array}{c}
0 \\
\frac{1}{\sqrt{2}}(\eta+\rho(x))
\end{array} \right) \eqno(3.13)
$$
Covariant derivative
$$
D_{\mu} \equiv \partial_{\mu} -ig_1 YB_{\mu}(x) -ig_2 T^a A_{\mu}^a(x)
\eqno(3.14)
$$
for the vacuum field $H_{vac}$
$$
D_{\mu}H_{vac} = (-ig_1 \frac{1}{2}B_{\mu} -ig_2 \frac{1}{2}\tau^a A_{\mu}^a)
\left(
\begin{array}{c}
0 \\
\frac{\eta}{\sqrt{2}}
\end{array} \right) =
$$
$$
~~~ \eqno(3.15)
$$
$$
= \frac{(-i)}{2\sqrt{2}}\eta \left(
\begin{array}{c}
\sqrt{2} g_2 W_{\mu}^- \\
-g_2 A_{\mu}^3 + g_1 B_{\mu}
\end{array} \right)
$$

The mass term for vector fields originates from $(D_{\mu}H)^+
D_{\mu}H$ term in the Lagrangian. It looks like
$$
{\cal L}_{mass} = \frac{1}{4}(g_2 \eta)^2 W_{\mu}^+ W_{\mu}^- +\frac{1}{8}
\eta^2(g_2 A_{\mu}^3 -g_1 B_{\mu})^2 \eqno(3.16)
$$

From this expression we conclude that the massive combination of $A_{\mu}^3$
and $B_{\mu}$ (i.e. $Z$-boson) is
$$
Z_{\mu} = \frac{1}{\sqrt{g_1^2 +g_2^2}}(g_2 A_{\mu}^3 -g_1 B_{\mu})
\eqno(3.17)
$$
or that
$$
tg\theta_W = g_1/g_2 \eqno(3.18)
$$
 From eq. (3.16) it follows that
$$
m_W = \frac{1}{2} g_2\eta \eqno(3.19)
$$
and
$$
m_W = m_Z \cos\theta_W \eqno(3.20)
$$

It is very interesting that $Z$ boson should be heavier than $W$ boson! After
spontaneous violation there still remains unbroken $U(1)$ symmetry that
corresponds to massless photon.

If we introduce electric charge $e$ as a coupling constant of the photon we
can relate $g_{1,2}$ with $e$ and $\cos\theta_W$. Let us rewrite interaction
of $A_{\mu}^3$ and $B_{\mu}$ as an interaction of $A_{\mu}$ and $Z_{\mu}$
fields:
$$
(-ig_2 T_3)A_{\mu}^3 -ig_1 YB_{\mu} \equiv (-i) \frac{g_2}{\cos\theta_W} [T_3
-\sin^2\theta_W Q] Z_{\mu} +(-i)(g_1\cos\theta_W) Q A_{\mu} \eqno(3.21)
$$
This is identically rewritten universal expression for covariant derivative.
So eq. (3.21) is applicable to the left and right fermions and to the Higgs
doublet.

From eq. (3.21) it follows immediately that
$$
e = g_1\cos\theta_W = g_2 \sin\theta_W  \eqno(3.22)
$$

We complete the description of bosonic sector of the SM.

\vspace{5mm}

{\bf 3.3. Weak interactions of leptons and quarks.}

\hspace*{9mm} {\bf Neutral Current. Request for new particles.}

\vspace{3mm}

Now we are ready to calculate the amplitude for the first physical process,
for the decay of $\mu\to e\nu\tilde{\nu}$. Charged currents Lagrangian for
leptons looks like
$$
\Delta{\cal L}_{Charged} = \frac{g_2}{2\sqrt{2}} W_{\mu}^+
[\bar{\nu}\gamma_{\mu}(1+\gamma_5) e + ...] \eqno(3.23)
$$
where the dots are for the similar terms with $\mu$ and $\tau$ leptons.
Feynman diagram for the $\mu$-decay is presented in Fig.1\\
\begin{center}
\begin{fmfgraph*}(100,40) \fmfpen{thin}
\fmfleft{i1,i2}\fmfright{o1,o2}
\fmf{fermion}{i1,v1,i2}
\fmf{photon,label=$W$,l.side=right}{v1,v2}
\fmf{fermion}{o1,v2,o2}
\fmflabel{$\mu$}{i1}
\fmflabel{$\nu$}{i2}
\fmflabel{$\nu$}{o1}
\fmflabel{$e$}{o2}
\end{fmfgraph*}
\end{center}

The amplitude for
the decay can be read from this diagram and it is equal to
$$
T(\mu\to e\nu\tilde{\nu}) =[\frac{g_2}{2\sqrt{2}}]^2 \frac{1}{m_W^2 -q^2}
(\bar{\nu}\gamma_{\alpha}(1+\gamma_5)\mu)
(\bar{e}\gamma_{\alpha}(1+\gamma_5)\nu) \eqno(3.24)
$$

The momentum transfer $q$ from muonic current to electronic current is of the
order of muonic mass $m_{\mu}$. So if $m_W \gg m_{\mu}$ the amplitude looks
like a point-like interaction in Fermi theory.
$$
T_{Fermi} = \frac{G_F}{\sqrt{2}} j_{\alpha}^e(j_{\alpha}^{\mu})^+
\eqno(3.25)
$$

Comparing these two presentations for the same amplitude we conclude that
$$
\frac{G_F}{\sqrt{2}} = \frac{g_2^2}{8m_W^2} \eqno(3.26)
$$

Taking into account eq. (3.19) for $m_W$ we also get that v.e.v. $\eta$ is
directly connected with $G_F$:
$$
\eta = [\sqrt{2}G_F]^{-1/2} = 246 \; \mbox{\rm GeV} \eqno(3.27)
$$
$$
G_F \equiv G_{\mu} = 1.16639(2) \cdot 10^{-5} \; \mbox{\rm GeV}^{-2}
$$

To fix remaining two fundamental parameters $g_1$ and $g_2$ we have to choose
two other physical observables measured with the best accuracy. The choice is
evident. They are the fine coupling constant $\alpha$
$$
\alpha^{-1} = \frac{4\pi}{e^2} = 137.035985(61) \eqno(3.28)
$$
and the mass of $Z$-boson
$$
m_Z = 91.187(2) \;\; \mbox{\rm GeV}  \eqno(3.29)
$$

To calculate $g_1$ and $g_2$ we first have to calculate the mixing angle
$\theta_W$ in terms of $G_F$, $\alpha$ and $m_Z$. It is not difficult
exercise to show that
$$
\sin^2\theta_W \cos^2\theta_W = \frac{\pi\alpha}{\sqrt{2}(G_F m_Z^2)} \;\; .
\eqno(3.30)
$$

{\it Exercise 1}: Derive eq. (3.30).

Substituting the values of the parameters from eqs. (3.27), (3.28) and (3.30)
we get
$$
\sin^2 \theta_W = 0.2120
$$
$$
g_1 = \frac{\sqrt{4\pi\alpha}}{\cos\theta_W} = 0.34 \eqno(3.31)
$$
$$
g_2 = \frac{\sqrt{4\pi\alpha}}{\sin\theta} = 0.66
$$

So we are ready for the first prediction in SM: we can calculate $m_W$
$$
(m_W)^{theor} = m_Z\cos\theta_W = 80.94 \;\; \mbox{\rm GeV} \eqno(3.32)
$$
that has to be compared with the current experimental value
$$
(m_W)^{exp} = 80.37(8) \;\; \mbox{\rm GeV} \eqno(3.33)
$$

The deviation from theoretical number is only 0.6\%, but
this tiny number is equal to $8\sigma$ deviation. To explain the huge
discrepancy we have to take into account radiative correction that have the
scale of the few per mill.

 The old 4-fermionic point-like theory is the effective theory for momentum
transfer much smaller than $m_W$. In this sense the SM is generalization of
the old theory. But SM also predicts the new phenomena that were unknown in V-A
theory. This is the neutral currents.

The effective 4-fermionic coupling of neutral currents is generated by $Z$
boson exchange.\\
\begin{center}
\begin{fmfgraph*}(40,80) \fmfpen{thin}
\fmfleft{i1,i2}\fmfright{o1,o2}
\fmf{plain}{i1,v1,o1}
\fmf{photon,label=$Z$,l.side=right}{v1,v2}
\fmf{plain}{o2,v2,i2}
\fmflabel{$\mu$}{i1}
\fmflabel{$e$}{i2}
\fmflabel{$\mu$}{o1}
\fmflabel{$e$}{o2}
\end{fmfgraph*}
\end{center}
 At small momentum transfer it is local
interaction with the coupling constant equal to $G_F \cos^2\theta_W$.

{\it Exercise 2}. Calculate the coupling constant for neutral currents.

Though this coupling is of the same order as $G_F$ by some reasons the
experimental search for neutral currents gave negative results for a long
time and only in 1973 experimental groups at CERN observed neutral currents
and provided the first experimental measurements of $\cos\theta_W$. This
measurement gave the possibility to calculate $m_W$ and $m_Z$ theoretically
(eqs. (3.19), (3.20)) with rather good accuracy. This estimate had been
extremely helpful for the experimental discovery of $W$ and $Z$ bosons.

Another great achievement of the SM was the request for new particles needed
for self-consistency of the theory. In 1970 the set of the known particles
included 2 generations of leptons
$$
\left(
\begin{array}{c}
\nu_e \\
e
\end{array}
\right)_L \; , \;\; \left(
\begin{array}{c}
\nu_{\mu} \\
\mu
\end{array}
\right)_L \;\; ; \;\; e_R \; , \mu_R  \eqno(3.34)
$$
and three quarks $u$, $d$ and $s$ that belong to the following $SU(2)_L
\times U(1)_Y$ representation
$$
\left(
\begin{array}{c}
u \\
d' = d\cos\theta_c +s \sin\theta_c
\end{array}
\right)_L \; , \;\; u_R \; , \;\; d_R \; , \;\; s_R  \eqno(3.35)
$$
where $\theta_c$ is the Cabibbo angle. First of all there was no symmetry
between quarks and leptons. No less important was the fact that for this 
set of quarks
$Z$ boson exchange produces flavour-changing $s\leftrightarrow d$ neutral
currents
$$
Z_{\mu}\bar{d}'_L \gamma_{\mu} d'_L \sim Z_{\mu}\left[(\bar{d}d)\cos^2\theta_c
+ (\bar{s}s)\sin^2\theta_c + \right.
$$
$$
\left. + \sin\theta_c
\cos\theta_c [\bar{d} s +\bar{s} d]\right] \eqno(3.36)
$$

This was absolutely forbidden by experimental data. To save the SM Glashow,
Illiopoulous and Maiani in 1970 introduced fourth $c$ quark and the new
$SU(2)_L$ doublet
$$
\left(
\begin{array}{c}
c \\
-d\sin\theta_c + s\cos\theta_c
\end{array}
\right)  \eqno(3.37)
$$

As a result flavour-changing neutral currents disappear and all neutral
currents become diagonal. This theoretical request for new particle was
satisfied by experimental discovery of $c$-quark in 1974.

\vspace{5mm}

{\bf 3.4. Quark masses. CKM matrix and CP-violation.}

\vspace{3mm}

In the Standard Model the standard mass term for the electron violates
$SU(2)_L$. Indeed this term
$$
m_e \bar{e} e = m_e[\bar{e}_R e_L +\bar{e}_L e_R ] \eqno(3.38)
$$
transforms like doublet instead of being invariant.

To preserve $SU(2)_L \times U(1)_Y$ symmetry we have to use Higgs mechanism
to generate the masses for fermions. For example Yukawa coupling of $L$, $e_R$ and $H$
is $SU(2)_L \times U(1)_Y$ invariant
$$
\Delta {\cal L} = f_e(\bar{L} e_R)H + h.c. =
$$
$$
= \frac{f_e}{\sqrt{2}}(\eta +\rho(x)) \bar{e} e = \eqno(3.39)
$$
$$
= m_e \bar{e} e + \frac{f_e}{\sqrt{2}} \rho(x) \bar{e} e
$$
where $\rho(x)$ is the field for physical Higgs in SM. From eq. (3.39) it
follows that Yukawa coupling is proportional to $m_e$
$$
f_e = \frac{\sqrt{2}}{\eta} m_e \simeq 3 \cdot 10^{-6}  \eqno(3.40)
$$

Notice that before this step the fields $e_L(x)$ and $e_R(x)$ were absolutely
different, i.e. they had different interaction with $W$ and $Z$. Yukawa
interaction unified this two Weyl spinors into one massive particle --
electron. To give the mass to down quarks we can use the same type of Yukawa
interaction
$$
\Delta{\cal L}_{m_d} = f_d(\bar{Q}_L d_R)H  \eqno(3.41)
$$
As for the mass of up quarks we need Higgs doublet with nonzero v.e.v. for up
component of doublet. At that moment we can introduce new Higgs doublet.
But in the case of $SU(2)$ group complex conjugated fields
$$
\tilde{H} = (-i\sigma_2) H^*   \eqno(3.42)
$$
also behave like a member of $SU(2)$ doublet. So we can use $\tilde{H}$ to
give mass to upper quark
$$
\Delta{\cal L}_m = f_d(\bar{Q}_L d_R) H + f_u(\bar{Q}_L u_R) \tilde{H}
\eqno(3.43)
$$
This is the solution of problem of fermion mass in the case of one generation. 
For more than one generation we have to take into account quark mixing.

\vspace{3mm}

{\bf Cabibbo-Kobayashi-Maskawa (CKM) matrix.}

For three generations of quarks the general Yukawa couplings produce general
non-diagonal mass $3\times 3$ matrix
$$
\Delta{\cal L} = (\bar{u}'_L)_i M_{ik}^{(u)} (u'_R)_k + (\bar{d}'_L)_i
M_{ik}^{(d)} (d'_R)_k + h.c.  \eqno(3.44)
$$
where $u'_i, d'_i$ are quarks that belong to $SU(2)$ doublet of $i$-th
generation $i=1,2,3$.

The matrices $M^{(u)}$ and $M^{(d)}$ can be diagonalized by use of left and
right unitary rotation, i.e.
$$
M^{(u)} = U_L^+ M_{diag}^{(u)} U_R
$$
$$
~~~ \eqno(3.45)
$$
$$
M^{(d)} = D_L^+ M_{diag}^{(d)} D_R
$$
where $M_{diag}^{(u,d)}$ are diagonal matrices. Substituting this expression
into eq. (3.44) we get that diagonal massive quark fields are
$$
u_R = U_R u'_R \;\; , \;\; u_L = U_L u'_L \;\; ;
$$
$$
~~~~ \eqno(3.46)
$$
$$
d_R = D_R d'_R \;\; , \;\; d_L = D_L d'_L \;\; ;
$$
It means that the members of the $SU(2)$ doublet are the mixture of different
massive fields. We already met such doublet in the case of 2 generations (see
eqs. (3.35) and (3.37)).

In terms of massive fields the charged currents look like
$$
j_{\mu}^+ = \bar{u}_L \gamma_{\mu} d'_R = \bar{u}_L(U_L D_L^+) \gamma_{\mu}
d_L  \eqno(3.47)
$$
The unitary $3\times 3$ matrix
$$
V = U_L D_L^+  \eqno(3.48)
$$
is the famous CKM matrix.

Due to unitarity of $U$ and $D$ matrices the neutral currents remain diagonal
$$
\bar{u}'_R u'_R = \bar{u}_R u_R \; , \;\;
\bar{u}'_L u'_L = \bar{u}_L u_L \; ,  \mbox{\rm etc.}
\eqno(3.49)
$$

You will have a course of lectures on CKM matrix by Y.Nir. I do not want to
interfere with his lectures but I'd like to mention one fundamental property of CKM. In
general case $n\times n$ unitary matrix can be represented as the product of
different orthogonal rotations dependent on $n_a$ angles
$$
n_a = \frac{1}{2} n(n-1)  \eqno(3.50)
$$
and of the $U(1)$ factors with $n_{ph}$ observable  phases 
$$
n_{ph} = \frac{1}{2}(n-1)(n-2)  \eqno(3.51)
$$
and of a number of nonobservable $U(1)$ factors that can be absorbed into
redefinition of quark's fields. So for three generation there is one
observable phase, i.e. charge currents contain complex couplings. This
immediately gives violation of CP-invariance. Till now it is unknown whether
this is the only source of CP-violation in the Nature. But in any case CKM
mechanism of CP-violation does exist.

\vspace{5mm}

{\bf 3.5. Subtle point. Triangle Anomaly.}

\vspace{3mm}

To have renormalizable theory of electroweak interactions it was absolutely
crucial to start from the gauge invariant theory where gauge bosons interact 
with conserved Noether currents. Spontaneous violation of symmetry does not 
spoil any symmetric relations between operators. They are exactly the same as 
in non-violent theory. The confusing notion of spontaneous violation means only nonlinear realization of the symmetry in the space of physical states.

In the SM we operate both with vector and axial currents. For any axial
currents
$$
j_{\mu}^A = \bar{\Psi} \gamma_{\mu}\gamma_5 \Psi
$$
$$
~~~~ \eqno(3.52)
$$
$$
\partial_{\mu} j_{\mu}^A = 2 i m \bar{\Psi}\gamma_5 \Psi
$$
So naively for massless fermion axial current is conserved. But what is true
in Classical Field Theory can be not true in Quantum Field Theory. Indeed
one-loop calculation of the divergence of axial current for electrons gives
instead of eq. (3.52)
$$
\left\{
\begin{array}{l}
\partial_{\mu} j_{\mu}^5 = 2 i m \bar{\Psi}_e \gamma_5 \Psi_e +
\frac{\alpha}{2\pi} F_{\mu\nu} \tilde{F}_{\mu\nu} \\
~~~ \\
\tilde{F}_{\mu\nu} = \frac{1}{2} \varepsilon_{\mu\nu\alpha\beta}
F_{\alpha\beta}
\end{array}
\right.  \eqno(3.53)
$$
The term $F\tilde{F}$ originates from matrix element of $\partial_{\mu}
j_{\mu}^5$ between vacuum and two-photon states.
\begin{center}
\begin{fmfgraph*}(100,80) \fmfpen{thin}
\fmfleft{i1,i2}\fmfright{o1}
\fmf{photon}{i1,v1}
\fmf{photon}{i2,v2}
\fmf{plain,tension=0.5}{o1,v1,v2,o1}
\fmflabel{$\gamma$}{i1}
\fmflabel{$\gamma$}{i2}
\fmfv{decor.shape=cross,decor.size=4thick}{o1}
\end{fmfgraph*}
\end{center}
So the axial current is not conserved even for $m\equiv 0$. Not any classical
symmetry can survive in Quantum Mechanics. This very interesting theoretical
phenomenon has special name -- triangle anomaly.

In the SM there are plenty of axial currents that interact with gauge fields.
Though fermions are massless (no mass terms in the Lagrangian) the anomaly
can destroy the conservation of Noether currents and this will kill
renormalizability. There is one possibility to save it. We see from eq.
(3.53) that anomaly depends only on the "charge" of particle that  is running
inside loop. So if the total gauge current has different pieces it can happen
that nonzero individual anomalies cancel each other for total current.

This cancellation imposes some restrictions on the charges of quarks
and leptons.  Let us check this possibility. We will calculate
the triangle matrix elements
between fields $A_{\mu}^i$ and $B_{\mu}$. There
are two crossing diagrams that contribute to anomalous interaction between 3
gauge fields. 

$$
\begin{array}{ll}
\begin{fmfgraph*}(100,80) \fmfpen{thin}
\fmfleft{i1,i2}\fmfright{o1}
\fmf{plain,tension=0.7}{v1,v2,v3,v1}
\fmf{photon}{i1,v1}
\fmf{photon}{i2,v2}

\fmf{photon}{o1,v3}
\end{fmfgraph*}
&
\begin{fmfgraph*}(100,80) \fmfpen{thin}
\fmfleft{i1,i2}\fmfright{o1}
\fmf{plain,tension=0.7}{v1,v2,v3,v1}
\fmf{phantom}{i1,v1}
\fmf{phantom}{i2,v2}
\fmf{photon}{o1,v3}
\fmffreeze
\fmf{photon}{i1,v2}
\fmf{photon,rubout}{i2,v1}


\end{fmfgraph*}
\end{array}
$$

Consider first the anomalous contribution of one
generation of matter. It is easy to see that:

1) ($A, A, A$) and ($A, B, B$) anomalies are automatically disappeared for
lepton doublet and for quark doublet separately.

2) ($B, A, A$) anomaly is disappeared if
$$
Q_e + 2Q_u + Q_d \equiv 0 \;\; ,  \eqno(3.54)
$$
i.e. quark contribution cancels lepton contribution only for this special
relation between charges. This relation means that hydrogen atom has to be
neutral!

It is very interesting that renormalizability of the SM takes place only if
the charge of proton is opposite to the charge of electron.

We can proceed further and consider other anomalies. At that moment we have
to make some statement about $\nu_R$. Suppose first that it does not exist at
all. In this case:

3) Cancellation of ($B, B, B$) anomaly takes place only if
$$
Q_e = -1 \; , \;\; Q_{\nu} =0 \; ; \;\; Q_u = \frac{2}{3} \; , \;\; Q_d =
-\frac{1}{3}  \eqno(3.55)
$$
(We suppose that QCD has $SU(3)_c$ symmetry.)

4) Cancellation of ($B \to$ gluon $+$ gluon) anomaly is automatic.

5) Cancellation of ($B \to$ graviton $+$ graviton) anomaly takes place only
for the charge sample eq. (3.55). So we are able to fix the relative charges
of leptons and quarks in this case.

If $\nu_R$ does exist anomalies 3) - 5) are disappeared automatically for any
charge of neutrino.

{\it Exercise IV.} Prove 1) - 5).

If we suppose that the new generations are the exact replica of the old one
(only masses are different, but the charges are the same) then we come to the
same conclusion for each generation. If we allow to change the charges from
generation to generation the restrictions on the choice of charges becomes
weaker. But in any case it is very interesting that renormalizability impose
restrictions on the property of matter fields.

\newpage

Lecture IV. {\bf Higgs, {\boldmath $W$} and {\boldmath $Z$}.}

\vspace{5mm}

Higgs boson $H$ is the only missing particle in the Standard Model. The
interaction of $H$ with gauge bosons $Z$, $W$ and
photon $\gamma$ is fixed by gauge
invariance of the SM. There is no freedom in this interaction.

As for the Yukawa coupling constants of $H$ with quarks and leptons they are
free parameters of the theory. Yukawa coupling constants determine the masses
of quarks and leptons, the mixing CKM angles and CP-odd phase. We do not
understand yet why these parameters of SM have given value and not
another one. So we have to take it from the experimental data.

A little can be said about Higgs particle. In this lecture we are going to
discuss this isoteric subject.

\vspace{5mm}

{\bf 4.1. Higgs Sector. Custodial symmetry of Higgs Potential.}

\vspace{3mm}

The Higgs potential in the SM

$$
V(H) = \frac{\lambda}{4}(H^+ H - \frac{\eta^2}{2})^2
\eqno(4.1)
$$
has been constructed to be $SU(2) \times U(1)$ invariant. Here
$H=\left(\stackrel{H^+}{_{H^0}} \right)$
is $SU(2)$ doublet with $Y_L =1/2$, $H^+ H$ is
singlet. So $SU(2) \times U(1)$ symmetry is absolutely evident.

In the unitary representation
$$
H(x) = exp [i\alpha_a(x) \frac{\tau_a}{2}] \left(
\begin{array}{c}
0 \\
\frac{1}{\sqrt{2}}(\eta + \rho(x))
\end{array}
\right)
\eqno(4.2)
$$

$SU(2) \times U(1)$ transformations change three Goldstone fields
$\alpha_a(x)$ and do not change the modulus $\rho(x)$. Potential $V(H)$ in
this representation depends only on the modulus $\rho(x)$:
$$
V(H) = V(\rho) = \frac{1}{2} (\frac{1}{2}\lambda \eta^2) \rho^2(x) +
\frac{1}{4} \lambda\eta\rho^3 + \frac{1}{16}\lambda\rho^4
\eqno(4.3)
$$

Potential (4.1) has minima at nonzero value of $H$. Quantization near
one of such minima
$$
<H>_{vac} = \frac{1}{\sqrt{2}} \left(
\begin{array}{c}
0 \\
\eta
\end{array} \right)
\eqno(4.4)
$$
spontaneously violates $SU(2) \times U(1)$ symmetry up to the $U(1)$ symmetry
that remains unbroken.

In the SM the v.e.v. $\eta$ is connected with Fermi coupling constant $G_F$
(see previous lecture) and is equal numerically to
$$
\eta = 246 \; \mbox{\rm GeV} \;\; .
$$

In the unitary representation potential $V(\rho)$ eq. (4.3) has 3
terms: quadratic, cubic and quartic ones. The quadratic term determines the
mass of Higgs boson
$$
m_H^2 = \frac{1}{2}\lambda\eta^2
\eqno(4.5)
$$
and cubic and quartic terms determine self-interaction of Higgs bosons.

In general theory the mass and the self-interaction are independent
parameters but not in the SM. Potential (4.1) depends only on two
parameters $\eta$ and $\lambda$. So the Higgs boson mass completely fix the
strength of self-interaction. For example the lower bound for $m_H$ from
LEP II experiment
$$
m_H \ga 90 \; \mbox{\rm GeV}
\eqno(4.6)
$$
can be immediately transformed into lower bound for self-interaction coupling
$$
\lambda = \frac{2m_H^2}{\eta^2} \ga 0.27
\eqno(4.7)
$$

For large $m_H$ the interaction of Higgs bosons becomes getting strong
$\lambda \ga 1$. In this case one should worry about large corrections to the
low-energy observables, say to the ratio $m_W /m_Z$.

Naively one can expect that for $\lambda \sim 1$ the corrections are of the
order of unity and that in this case the SM looses its predictive power.
Fortunately this naive expectation is not correct and corrections to
electroweak observables at low-energy (i.e. to LEP observables) are small
even for heavy Higgs bosons. The reason is very subtle and beautiful. There
is hidden symmetry of Higgs that preserves low-energy observables from large
corrections. This is custodial $SU(2) \times SU(2)$ symmetry. Potential
(4.1) that looks like $SU(2) \times U(1)$ invariant possesses larger
$SU(2) \times SU(2)$ symmetry.

To find this symmetry it is convenient to rewrite (4.1) using new
representation for doublet $H$:
$$
H = \left(
\begin{array}{c}
H^+ \\
H^0
\end{array}
\right) = (\pi_0 + i\pi_a \tau_a) \left(
\begin{array}{c}
0 \\
1
\end{array}
\right) = \left(
\begin{array}{c}
\pi_2 + i\pi_1 \\
\pi_0 -i\pi_3
\end{array}
\right)
\eqno(4.8)
$$
where $\pi_0$ and $\pi_a$ are real scalar fields. At first sight we have done
nothing but changed a little bit the notation for $H^+$ and $H^0$. It is
true, but in this new notation $H^+ H$ looks like a scalar product of
four-vector $\pi_{\mu} = (\pi_0, \pi_a)$. Indeed
$$
H^+ H = (0,1)[\pi_0 -i\vec{\pi}\vec{\tau}] [\pi_0 + i\vec{\pi}\vec{\tau}]
\left(
\begin{array}{c}
0 \\
1
\end{array}
\right)
=\pi_0^2 +\vec{\pi}^2 = \pi_{\mu}\pi_{\mu} \;\; .
\eqno(4.9)
$$

The quadratic form $\pi_{\mu}\pi_{\mu} = \pi_0^2 +\vec{\pi}^2$ is invariant
with respect to $O(4)$ orthogonal transformation and $O(4)$ group can be
presented as
$$
O(4) = SU(2) \times SU(2) \;\; .
\eqno(4.10)
$$

So for one Higgs doublet $H$ any $SU(2) \times U(1)$ invariant potential
$V(H^+ H)$ is $SU(2) \times SU(2)$ invariant as well.

After spontaneous violation of symmetry the field $\pi_0$ gets nonzero v.e.v.
and surviving symmetry is $O(3)$ rotational symmetry between different
components of $\pi_i$ or as a rotational symmetry between Goldstone fields
$\alpha_i$:
$$
\alpha_i(x) \to \alpha'_i(x) = O_{ij} \alpha_j(x)
\eqno(4.11)
$$
This $O(3)$ symmetry is custodial symmetry of Higgs potential.

The immediate consequence of the existence of the custodial symmetry is
that the relation between $m_W$ and $m_Z$
$$
m_W /m_Z = \cos\theta
\eqno(4.12)
$$
is correct to any order of $\lambda$. On other words there are no corrections
of the order of $\lambda^2, \lambda^3$ etc. that could charge the prediction
(4.12) by order of magnitude for $\lambda \ga 1$. Why?

If we start the calculation of the corrections of the order of $\lambda^2,
\lambda^3$ etc. to the effective action we immediately find such large
corrections to the v.e.v. $\eta$ and to the wave function renormalization $Z$
of Goldstone fields $\alpha_i$. (Wave function renormalization of gauge
fields and gauge coupling constants are proportional to the square of gauge
coupling constant and is considered as the small corrections.) Due to $O(3)$
custodial symmetry (4.11) of the selfinteraction the renormalization of
the Goldstone fields should be the same for each of the components
$\alpha_i(x)$, so that renormalized fields $\alpha^R(x)$ are transformed as a
vector under $O(3)$ custodial symmetry
$$
\vec{\alpha}^R(x) = Z^{1/2} \vec{\alpha}^B(x)
\eqno(4.13)
$$
where $\vec{\alpha}^B$ are bare Goldstone fields. This is the central point.
Indeed since $\vec{\alpha}^B(x)$ is a
$O(3)$ vector it can be eaten by gauge
transformation. In this gauge (unitary gauge) the renormalized effective
action looks like the bare action that we considered in section 3.1.
The only
difference is that the Higgs vertices, the Higgs field $\rho(x)$ and
v.e.v. $\eta$ are renormalized by self-interaction
in the order of $\lambda^1, \lambda^2$ ... So
in this approximation we get that
$$
m_W^R = (\pi \alpha_W)\eta^R
$$
$$
m_Z^R = (\pi \alpha_Z)\eta^R
\eqno(4.14)
$$
(see eqs. (3.19) and (3.20)) and that
$$
m_W^R/m_Z^R = \alpha_W/\alpha_Z = \cos\theta_W
$$

This relation is valid in any order of $\lambda$!

If we switch on the coupling constants $g_2$ and $g_1$ and consider the
corrections of the order of $\alpha$, $\alpha\lambda$, $\alpha\lambda^2$,
...  the
custodial $SU(2)\times SU(2)$ symmetry will be broken and we can expect the
correction of the same order $\alpha_1$, $\alpha\lambda$, $\alpha\lambda^2$,
... to the ratio $m_W/m_Z$ and to other observables. These
corrections do exist. They are of the order of two-loop electroweak
corrections if the Higgs boson mass is not much larger than $m_{W,Z}$
$$
\alpha_W \cdot \lambda \sim \alpha_W^2 \left(\frac{m_H}{m_Z}\right)^2
$$
These small corrections will be studied in the next lecture.

\vspace{5mm}

{\bf 4.2. Unitary Bound on Higgs mass.}

\vspace{3mm}

Coupling constant $\lambda$ of selfinteraction and the mass $m_H^2$ are
independent parameters in the general theory but in the SM they are
proportional to each other
$$
m_H = \sqrt{\frac{\lambda}{2}} \cdot \eta \simeq \sqrt{\frac{\lambda}{2}}
(246 \; \mbox{\rm GeV})
\eqno(4.15)
$$

If $m_H \gg m_{W,Z}$ the $SU(2) \times U(1)$ gauge theory looks like the
old-fashioned theory of massive vector particles that is nonrenormalizable
one. In nonrenormalizable theories scattering amplitudes grow with energy and
violate unitary bounds at some value of energy. Consider for example the
scattering of $W$ boson on $Z$ boson. There are three types of diagrams that
contribute into this process. 

$$
\begin{array}{ccc}
\begin{fmfgraph*}(60,60) \fmfpen{thin}
\fmfleft{i1,i2}\fmfright{o1,o2}
\fmf{photon}{i1,o2}
\fmf{photon}{i2,o1}
\fmflabel{$Z$}{i1}
\fmflabel{$Z$}{o1}
\fmflabel{$W$}{i2}
\fmflabel{$W$}{o2}
\end{fmfgraph*}
\;\;\;\;\;\;&
\begin{fmfgraph*}(80,60) \fmfpen{thin}
\fmfleft{i1,i2}\fmfright{o1,o2}
\fmf{photon}{i1,v1,i2}
\fmf{photon}{o2,v2,o1}
\fmf{photon,label=$W$,l.side=left}{v1,v2}
\fmflabel{$Z$}{i1}
\fmflabel{$Z$}{o1}
\fmflabel{$W$}{i2}
\fmflabel{$W$}{o2}
\end{fmfgraph*}
\;\;\;\;\;\;&
\begin{fmfgraph*}(80,60) \fmfpen{thin}
\fmfleft{i1,i2}\fmfright{o1,o2}
\fmf{photon}{i1,v1,o1}
\fmf{photon}{i2,v2,o2}
\fmf{dashes,label=$H$,l.side=left}{v1,v2}
\fmflabel{$Z$}{i1}
\fmflabel{$Z$}{o1}
\fmflabel{$W$}{i2}
\fmflabel{$W$}{o2}
\end{fmfgraph*}\\
\\
1a\;\;\;\;\;\;& 1b\;\;\;\;\;\;& 1c
\end{array}
$$

It is clear that for $m_H \to
\infty$ the diagrams (1c) goes to zero as $1/m_H^2$ and the remaining
two diagrams (1a,b) are the same as in the theory of massive $W$ and $Z$
without Higgs mechanism.

We are going to demonstrate that these two amplitudes  for longitudinally
polarized $W_L$ and $Z_L$ linearly grow with
$s$, where $s$ is square of the energy in the c.m.f.
Indeed any polarization vector of massive vector
particle satisfies the condition
$$
\varepsilon_{\mu}^V(p) p_{\mu} = 0
\eqno(4.16)
$$
For momenta $p_{\mu} =(\varepsilon, 0, 0, p)$ the solution of eq.
(4.16) with longitudinal components is
$$
\varepsilon_{\mu}^V (p) = \frac{1}{m_V}(p, 0, 0, \varepsilon) \simeq
$$
$$
\simeq  \frac{P_{\mu}}{m_V} +0(\frac{m_W}{\varepsilon}) \;\; ,
\eqno(4.17)
$$
i.e. longitudinal component grows with energy. So any of the
diagrams 1a,b is of the order of
$$
Amp \sim (\frac{p_1}{m_W})(\frac{p_2}{m_Z})(\frac{p_3}{m_W})
(\frac{p_4}{m_Z}) \sim \frac{s^2}{m_W^2 m_Z^2}
\eqno(4.18)
$$

So in general theory of massive vector particle the scattering amplitude $W_L
Z_L \to W_L Z_L$ grows quadratically with $s$. In the loops this dependence
transforms into divergence of the loop integrals and this is
why the theory of massive particle in general is nonrenormalizable.

But SM is a very special theory because it is gauge invariant. The sum of the
diagram 1a and 1b (and the diagram 1c itself) is  gauge invariant. It means
that gauge boson interacts with conserved Noether  currents. Consider
for example the interaction of incoming $W$:
$$
T_{1a} +T_{1b} =g_W \varepsilon_{\mu}^W(p_1) \;
<Z, W|J_{\mu}^W|Z>
\eqno(4.19)
$$
where $<Z, W|$, $<Z|$ represent the physical states of
remaining particles. Since the $J_{\mu}^W$ is conserved current any matrix
element of $J_{\mu}^W$ should be transversal
$$
(p_1)_{\mu} < |J_{\mu}^W(p_1)| > \equiv 0
\eqno(4.20)
$$

It means that in the case of the theory of massive particles with
gauge-invariant interaction the sum of two diagrams 1a and 1b eq.
(4.19) should be transversal with respect to $W$-boson momenta
$(p_1)_{\mu}$. If so the most dangerous term $\sim p^1_{\mu}/m_W$ in
polarization vector $\varepsilon_{\mu}^W$ gives zero contribution and the
remaining terms $\sim m_W/\varepsilon$ gives
$$
(T_{1a} +T_{1b}) \sim (\frac{m_W}{E})<Z|J_{\mu}|ZW> \sim (\frac{m}{E})
(\frac{E}{m})^3 \sim \frac{s}{m^2}
\eqno(4.21)
$$

The natural idea now is to kill in the same way the dangerous terms for
polarization vector of two $W$-bosons and to prove that amplitude does
not grow with energy. Unfortunately our trick
works only once. The interaction of the $W$-bosons with corresponding
conserved currents look like
$$
T_{1a} +T_{1b} = g_W^2 \varepsilon_{\mu}^W(p_1)\varepsilon_{\nu}^W(p_3)
\int d^4 x_1 d^4 x_2 e^{-i p_1 x_1} e^{ip_3 x_3} \times
$$
$$
\times <Z|T\{J_{\mu}^{W^+}(x_1) J_{\nu}^{W^-}(x_3)\} |Z>
\eqno(4.22)
$$
where $T\{\}$ is the $T$-product of two currents. Just because currents in
non-abelian theory do not commute the $T$-product does not allow to cancel
$p_{1\mu}$ and $p_{3\nu}$ -- dangerous terms  simultaneously.

{\it Exercise 1.} Prove this technical statement.

So we conclude that amplitudes without Higgs exchange grows linearly
with $s$. Unitarity of the scattering amplitudes says
that the any partial amplitude is less than 1. So sooner or later
this amplitude has to violate unitarity bound. The
value of energy $\sqrt{s_0}$ at which one of the partial waves intersects
unitary bounds varies both with angular momenta $l$ and from one process to
another one. We are not going to repeat the detailed calculations made in
literature. The result of these calculations gives
$$
\sqrt{s_0} \sim 700 \; {\rm GeV}
\eqno(4.23)
$$

To stop the growth of the amplitude we have to switch on the diagrams with
Higgs exchange. If we do it too late unitarity will be violated already. So
to prevent the violation of unitarity bound
$$
m_H \la 700 \; {\rm GeV} \;\; .
\eqno(4.24)
$$

What is wrong with heavy Higgs? Actually the violation of unitarity that we
found for tree diagrams means that the coupling constant becomes strong and
that we can't restrict ourselves by calculation of tree diagrams, we have to
include all loops and to calculate the whole infinite series in $\lambda$. We
do not know how to deal with strong interaction. So the bounds (4.22),
(4.23) and (4.24)
say that for $m_H <700$ GeV the theory can be treated perturbatively and for
heavier Higgs we have a theory of strong interaction and we can't calculate
$WZ \to WZ$ scattering theoretically. It is interesting to note that the
attempts to work in strong coupling regime (e.g. using computer simulation
for the Higgs theory on the lattice) demonstrate that the bound (4.23)
and (4.24)
is very stable, i.e. it takes place even for strong interacting Higgs.

\vspace{5mm}

{\bf 4.3. Effective potential. Stability of the Universe and Bounds}

\hspace*{9mm}{\bf on {\boldmath $m_H$}.}

\vspace{3mm}

The Higgs potential in the SM
$$
V_{\rm cl}(H) = \frac{\lambda}{4} H^4 -\frac{\mu^2}{2}H^2
\eqno(4.25)
$$
has minima that corresponds to nonzero v.e.v. of field $H$:
$<H>_{\rm vac} = \eta$. Loop corrections change self-interactions of
Higgs particles.

$$
\begin{array}{ccc}
\begin{fmfgraph*}(80,60) \fmfpen{thin}
\fmfleft{i1,i2,i3,i4}
\fmfright{o1,o2,o3,o4}
\fmf{plain}{i2,v1,i3}
\fmf{plain}{o2,v2,o3}
\fmf{plain,right,label=$H$,l.side=right,tension=0.5}{v1,v2}
\fmf{plain,right,tension=0.5}{v2,v1}
\end{fmfgraph*}\;\;\;\;\;\;&
\begin{fmfgraph*}(80,60) \fmfpen{thin}
\fmfleft{i1,i2,i3,i4}
\fmfright{o1,o2,o3,o4}
\fmf{plain}{i2,v1,i3}
\fmf{plain}{o2,v2,o3}
\fmf{fermion,right,label=$F$,l.side=right,tension=0.5}{v1,v2}
\fmf{fermion,right,tension=0.5}{v2,v1}
\end{fmfgraph*}\;\;\;\;\;\;&
\begin{fmfgraph*}(80,60) \fmfpen{thin}
\fmfleft{i1,i2,i3,i4}
\fmfright{o1,o2,o3,o4}
\fmf{plain}{i2,v1,i3}
\fmf{plain}{o2,v2,o3}
\fmf{photon,right,label=$W,, Z,, \gamma$,l.side=right,tension=0.5}{v1,v2}
\fmf{photon,right,tension=0.5}{v2,v1}
\end{fmfgraph*}
\end{array}
$$

The effective potential that takes
into account loops corrections was calculated by Coleman and Weinberg in
1973. In one-loop approximation it looks like
$$
V_{\rm eff}(H) - V_{\rm cl}(H) =\frac{1}{64\pi^2} \left\{ \frac{m_H^4 +6m_W^4 +3m_Z^4}{\eta^4} -
\frac{12m_t^4}{\eta^4} \right\} H^4 \ln \frac{H^2}{M^2}
\eqno(4.26)
$$
where we have neglected by small contributions from fermions other than
$t$-quark. Note that due to Fermi-Dirac statistic the contribution of
fermion loops has opposite sign in compared to the bosonic loops.

 It is clear that corrections (4.26) become more important than the main
classical potential (4.25) for very large field
$H$ when $\left(\frac{m}{\eta}\right)^4\ln H \ga
\left(\frac{m_H}{\eta}\right)^2$.

In one-loop approximation we get that the correction (4.26) has negative sign
if $m_H < \sqrt[4]{12} m_t$. For this case the effective potential has no
ground state(see fig. 2). 
\begin{figure}
\begin{center}
\epsfbox{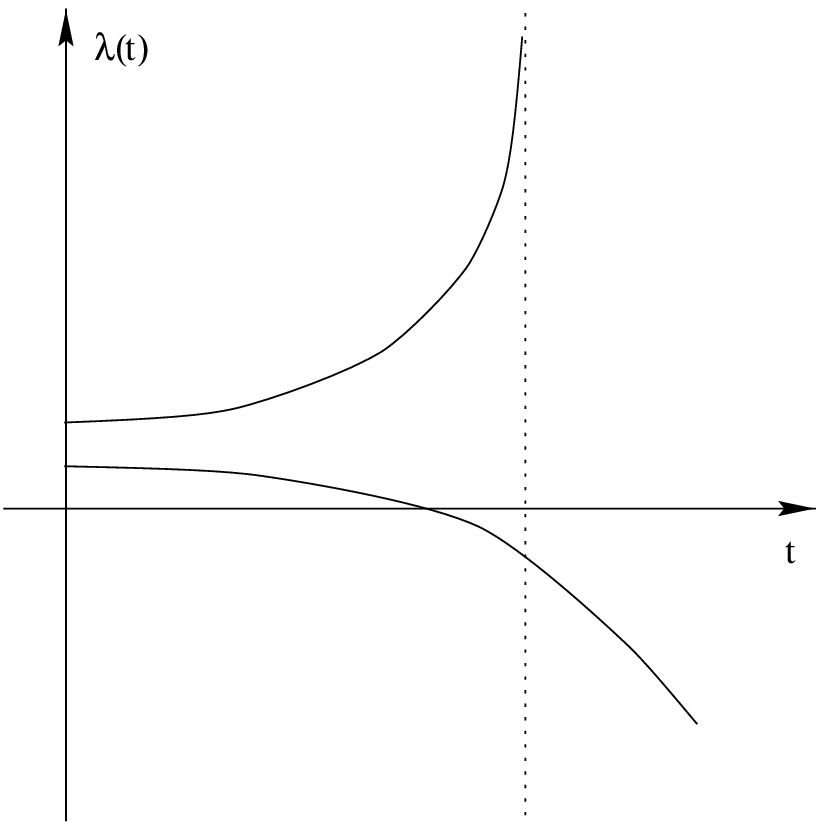}
\caption{}
\end{center}
\end{figure}

So even if our system  was located first in the local
minima at $< H > = \eta$ it will decay at $t\to \infty$ and the average
value of field H will run to infinity. We know that nothing like that
has happened with our Universe that is near $10^{10}$ years old.
So the stability of the Universe imposes strong constraints on the masses of
top and Higgs particles.

 We should improve a little our one-loop formula (eq. (4.26)). For large $H$
one-loop logarithmic corrections $\lambda\ln H$ and $\alpha_W \ln H$ are of the same order as tree terms, two-loop double-logarithmic terms are of the
order of one-loop terms etc. So all these logarithmic terms should be taken
into account. Fortunately this technical problem is not very difficult-
renormalization group techniques help to sum up such corrections.
The result is
$$
V^{\rm eff}(H) = -\frac{1}{2}\mu^2(t)H^2(t)
+\frac{1}{4}\lambda(t)H^4(t)
\eqno(4.27)
$$
where $\lambda(t)$ and $\mu(t)$ are running parameters and $t =\ln
H/\eta$. For small value of t (i.e. for small value of field $H$ ) the
running parameters do not run far away from their classical values and
the effective potential is equal to the classical one with the accuracy
of small radiative corrections.
 For large $H$ we can forget about $\mu^2 H^2$ and the whole
dynamics at large $H$ is governed by running coupling constant
$\lambda(t)$. There are different contributions into differential
equations for running $\lambda(t)$, coming from the loops with top quarks,
vector bosons and Higgs boson itself. If the top quark contribution dominates,
i.e. the higgs coupling to top (i.e. the top mass) is large, $\lambda(t)$
changes sign (see fig. 3) and the vacuum becomes unstable. This is
reformulation of the phenomena that we had at one loop level.

  If the Higgs selfinteraction dominates, i.e. the Higgs mass is large,
then the evolution of $\lambda(t)$ is similar to the evolution
of coupling in the $H^4$ theory without other fields.
It is known that in this case the behavior of $\lambda(t)$ is
$$
\lambda(t) = \frac{\lambda(t_0)}{1-b\lambda(t_0)\ln \frac{t}{t_0}}
\eqno(4.28)
$$
and running coupling goes to infinity at  some finite value of $H$ (see
fig. 3).

\begin{figure}
\begin{center}
\epsfbox{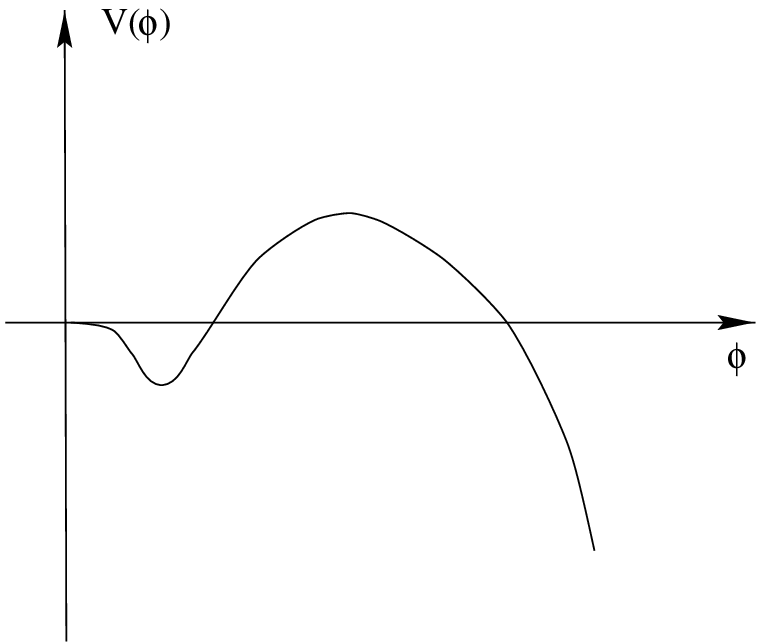}
\caption{}
\end{center}
\end{figure}

$$
H = \Lambda \;\; .
$$

This is the Landau pole in the running coupling constant. When initial
condition $\lambda(t_0)$ (i.e. the value of Higgs mass ) increases the value
of Landau pole goes done. If we substitute this running coupling constant
into eq.(4.27) we get that the effective potential runs to infinity at
this value of $H$ as well. This is some new phenomena.

Such singular behavior of the coupling constant is unacceptable from the
physical point of view. Indeed for any \underline{finite} value of the bare
coupling constant $\lambda^B$ ($\lambda^B$ is equal to the running coupling $\lambda(t)$ at the cut-off $\Lambda$) we get that renormalized coupling constant 
(i.e. $\lambda(t)$ at low value of $t$) is equal to \underline{zero}. It means 
that at low energy we get trivial free theory. This pathological theory seems 
to be unphysical.

There are two possibilities to improve bad behavior of $\lambda(t)$. The
first one is so to say dynamical. For $\lambda \sim 1$ the multiloop
corrections and nonperturbative corrections change $\lambda(t)$ at large $t$
so that Landau pole disappears. This is the possible solution of the problem
in the strong coupling regime.

Another solution can be reached in perturbation theory if some new physics
(i.e. new interactions and new particles) contribute into $\lambda(t)$ at scale
near $\Lambda$ so that pole disappears.

If we believe that there are no new physics up to some scale (or that the
theory can be treated perturbatively up to this scale) we have to push
the position of Landau pole
$\Lambda$ (calculated in one-loop approximation eq. (4.28)) to higher scale.
This impose upper bound on the value of Higgs mass. So we have bounded $m_H$ 
both from above and from below. This remarkable line of reasoning was invented
 by Cabibbo et al. in 1979.

There are different choices for the parameter $\Lambda$. For example $\Lambda$ can be of the order of Planck scale
$$
\Lambda \sim \Lambda_{Pl} = 10^{19} \; \mbox{\rm GeV} \;\; ,
$$
or of the order of Grand Unification Scale
$$
\Lambda \sim \Lambda_{GUT} = 10^{15} \; \mbox{\rm GeV} \;\; ,
$$
or of the scale of the energy of the accelerator of the next generation
$$
\Lambda \sim 10^3 - 10^5 \; \mbox{\rm GeV} \;\; .
$$

We have to keep in mind all these possibilities.
It is  evident that for the strongest assumption that new physics
does not appear up to the Planck scale we should get the strongest 
upper bound for $m_H$.

To derive more quantitative results we have to solve differential equation
for the running coupling constant $\lambda(t)$. The renormalization of
$\lambda(t)$ depends on self-interaction coupling, on gauge
coupling and on Yukawa coupling constants. So we have to solve the whole system
of coupled differential equations. This can be done numerically with the help
of computer. The result of calculation for $\Lambda = 10^{15}$ GeV is
presented in the fig. ~~~. 

\begin{figure}
\begin{center}
\epsfbox{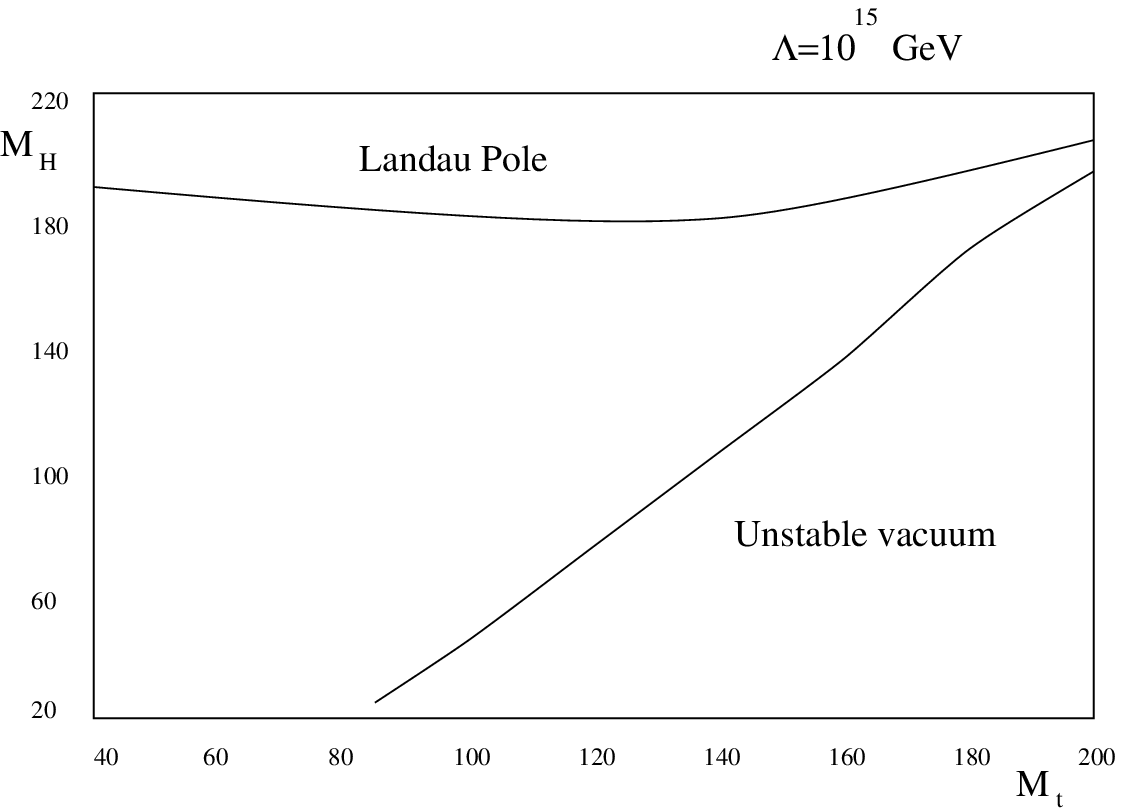}
\caption{}
\end{center}
\end{figure}

This is so to say the phase diagram in the plane
$m_t$ and $m_H$. Allowed region is located between two curves, the lower region
corresponds to unstable vacuum and for the parameters in the upper region Landau pole appears at the scale lower than $\Lambda = 10^{15}$ GeV. For experimental
value of $m_t \simeq 175$ GeV the allowed region for $m_H$ is very strong
$$
170 \; \mbox{\rm GeV} < m_H < 190 \; \mbox{\rm GeV} \;\; 
\;\; (\Lambda = 10^{15} \; \mbox{\rm GeV}).
$$
For $\Lambda \simeq 10^5$ GeV the upper bound is much weaker.

\vspace{5mm}

{\bf 4.4. Basic properties of $W$ and $Z$ bosons.}

\vspace{3mm}

The interactions of gauge bosons with quarks, leptons, Higgs bosons and with
gauge bosons itself are determined by the principle of gauge invariance. In
this section we are going to calculate the partial widths of $W$ and $Z$
boson.

According to eq. (3.21) we write the amplitude of the $Z$-boson decay into
fermion-antifermion pair $\bar{f}f$ in the form
$$
T(Z \to \bar{f}f) = \frac{e}{2sc}Z_{\alpha}\bar{\Psi}_f (g_V \gamma_{\alpha}
+g_A\gamma_{\alpha}\gamma_5)\Psi_f \;\; ,
\eqno(4.29)
$$

where
$$
g_V = T_3^f - 2s^2 Q^f
$$
$$
g_A = T_3^f
$$
$$
and
$$
$$
s = \sin\theta \; , \;\; c = \cos\theta.
$$

With good accuracy we can neglect the masses of fermions and derive
the universal formula for any partial width of $Z$ boson
$$
\Gamma_Z\to\bar{f}f) =\Gamma_0[g_V^2 +g_A^2] \;\; ,
\eqno(4.30)
$$
where $\Gamma_0$ is the standard width:
$$
\Gamma_0 = \frac{G_F m_Z^3}{6\sqrt{2}\pi} = 332 \; \mbox{\rm MeV}
\eqno(4.31)
$$
For neutrino we have
$$
g_V = g_A = \frac{1}{2} \;\; ,
$$
so that
$$
\Gamma_{inv} = \sum_i \Gamma(Z\to \nu_i \tilde{\nu}_i) = 3 \cdot
\Gamma_0(\frac{1}{4} +\frac{1}{4}) = 498 \; \mbox{\rm MeV} \;\; .
\eqno(4.32)
$$
This is theoretical prediction for the decay of $Z$ boson into invisible
modes.

For the decays to any of the pairs of charged leptons we have
$$
g_A = -\frac{1}{2}
$$
$$
g_V =(-\frac{1}{2})(1-4s^2)
$$
and
$$
\Gamma_l = \Gamma(Z\to l\bar{l}) = \frac{1}{4}[1+(1-4s^2)^2]\Gamma_0 \simeq
83.47 \; \mbox{\rm MeV}
\eqno(4.33)
$$
Note that for $s^2 \simeq 0.2320$ the vector  coupling $g_V$ for charged
lepton is a very small number!

For the decays into any of five pairs of quarks we have
$$
\Gamma_q = \Gamma(Z\to \bar{q}q) = 3\Gamma_0[g_A^2 R_A
+g_V^2 R_V]
\eqno(4.34)
$$
where the factor $3$ is due to three color of each quarks and the factors
$R_{A,V}$ are due to exchange of gluons between quarks in the final state. In
the first order of strong coupling $\hat{\alpha}_s = \alpha(q^2 =M_Z^2)$ they
are equal
$$
R_A = R_V = 1+\frac{\hat{\alpha}_s}{\pi}
\eqno(4.35)
$$
As a result
$$
\Gamma_h = \Gamma(Z\to \mbox{\rm hadrons}) = 3\Gamma_0 \left\{ 2[\frac{1}{4}
+(\frac{1}{2} -\frac{4}{3}s^2)^2] + \right.
$$
$$
\left.+3[\frac{1}{4}+(\frac{1}{2} -\frac{2}{3}s^2)^2]
\right\}(1+\frac{\alpha_s}{\pi}) = 1676(1+\frac{\hat{\alpha}_s}{\pi}) \;
\mbox{\rm MeV}
\eqno(4.36)
$$
The numerical value of $\hat{\alpha}_s$ found from deep-inelastic processes
is of the order of $\hat{\alpha}_s \simeq 0.119$.
So the corrections due to strong interaction in the final state are of the
order of 4\%. 

For the total width we have
$$
\Gamma_Z^{th} = \Gamma_{inv} +3\Gamma_l +\Gamma_h \simeq 2488.5 \mbox{\rm
MeV}
\eqno(4.37)
$$
that should be compared with the experimental LEP and SLC value
$$
\Gamma_Z^{ex} = 2493.9 \pm 2.4 \; \mbox{\rm MeV} \;\; .
\eqno(4.38)
$$
Though the agreement between $\Gamma_Z^{th}$, calculated in the tree
approximation, and $\Gamma_Z^{ex}$ is very good the experimental accuracy 
is better. So we
have to improve the accuracy of the theoretical calculation and to take into
account the radiative corrections (see lecture V).

Consider now the decays of $W$ boson. According to eq. (3.23) the amplitude
of $W^+$ decay into $\tilde{\nu}_l$l can be written in the form
$$
T(W\to\tilde{\nu}_l l) =\frac{e}{2\sqrt{2}s}
W_{\alpha}(\bar{l}\gamma_{\alpha}(1+\gamma_5)\nu)
\eqno(4.39)
$$
Neglecting leptonic mass we get
$$
\Gamma(W\to l\tilde{\nu}) = \frac{1}{6\sqrt{2}\pi}G_{\mu}m_W^3 \simeq 226 \;
\mbox{\rm MeV} \;\; .
\eqno(4.40)
$$
Consider now the hadronic decays of $W$-boson. The decay width into top quark
is equal to zero since top quark is heavier than $W$-boson. With good
accuracy we can neglect the masses of other quark. In this approximation
$$
\Gamma(W\to \mbox{\rm hadrons}) = 3 \cdot 2\Gamma(W\to e\tilde{\nu})
\eqno(4.41)
$$
where factor $3$ is due to color of each quark and factor $2$ is due to two
quark channels of $W$ decay.

\vspace{3mm}

{\it Exercise:} Prove that in massless approximation hadronic width eq. (4.41)
does not depend on the CKM angles.

In this approximation the theoretical prediction for the total width is
$$
\Gamma_W^{th} \simeq 9\Gamma(W\to\tilde{\nu}_l l) \simeq  2.03 \; \mbox{\rm GeV}
\eqno(4.42)
$$
that has to be compared with the experimental value
$$
\Gamma_W^{exp} \simeq 2.06 \pm 0.06 \; \mbox{\rm GeV}\;\; .
\eqno(4.43)
$$

As it was expected the agreement is of the order of percent. The experimental
accuracy is of the same order.

\end{fmffile}
\newpage
Lecture V. {\bf Radiative Correction in SM.} \\
\hspace*{9mm} {\bf Hunting for virtual $t$-quark.}

\vspace{5mm}

{\bf 5.0. $Z$-physics at LEP and SLC.}

\vspace{3mm}

To test the predictions of the SM the huge "factories" of $Z$-bosons ($e^+ e^-$ colliders) were constructed of CERN (LEPI) and at SLAC (SLC). Electrons and positrons in thus colliders collide at the centre of mass energy equal to the $Z$-boson mass. The reactions that are studied can be presented in the form 

$$ e^+e^- \rightarrow Z \rightarrow \bar{f}f$$
where

$$ \bar{f}f = \left\{ 
               \begin{array}{lr}
               \nu \tilde{\nu} &~~~~~~~invisible~modes \\
                l \bar{l}     &~~~~~~~charged~leptons \\
                q \bar{q}     &~~~~~~~hadrons.
               \end{array}
               \right. $$ 

The LEPI was terminated in the fall 1995 in order to give place to LEP II. Four groups --- collaborations ALEPH, DELPHI, L3, OPAL ---
have collected near $2\cdot 10^7$ $Z$-bosons.
The SLC has worked from the fall 1989 till the fall 1998. The SLD detector recorded near $5\cdot10^5$ $Z$-bosons. 
The electrons in the SLC were longitudinally polarized so that SLD group could provide very high precision data having relatively low statistics. 
More than two thousand experimentalists and engineers and hundreds of
theorists participated in this unique project -- one of the largest project in the history of physics. 

Near the dozen of independent observables were measured with fantastic precision of the order of $10^{-3}$ ($10^{-5}$ for the case of $Z$-boson mass).  The scale of the radiative corrections in the SM is of the order of 
$\alpha_{W,Z}/\pi \sim 10^{-2} - 10^{-3}$. Therefore LEP-I and  SLD data provide precision test of the SM as a renormalizable field theory, i.e.
with loops included.

The theoretical study of electroweak corrections in SM started in 1970's and was elaborated by a number of theoretical groups. Near $10^2$ theorists had published near $2\cdot 10^3$ papers on radiative corrections (see e.g.ref.[2]). This study has been summarized in Yellow CERN Reports of Working Groups on precision calculations for the Z resonance in 1995 [3]. The deviations in theoretical calculations of different groups are by the order of magnitude smaller than the experimental uncertainties.

By comparing the $\Gamma_{invisible}$ with theoretical predictions for neutrino decays the result of fundamental importance was established -- the sequence of the generations with light neutrino is completed with number of generation \\
$N_f = 3$. 

The best fit of the most recent data submitted to Vancouver (1998) conference is presented at Table I.
\begin{figure}
\epsfysize=500pt
\begin{center}
\epsfbox{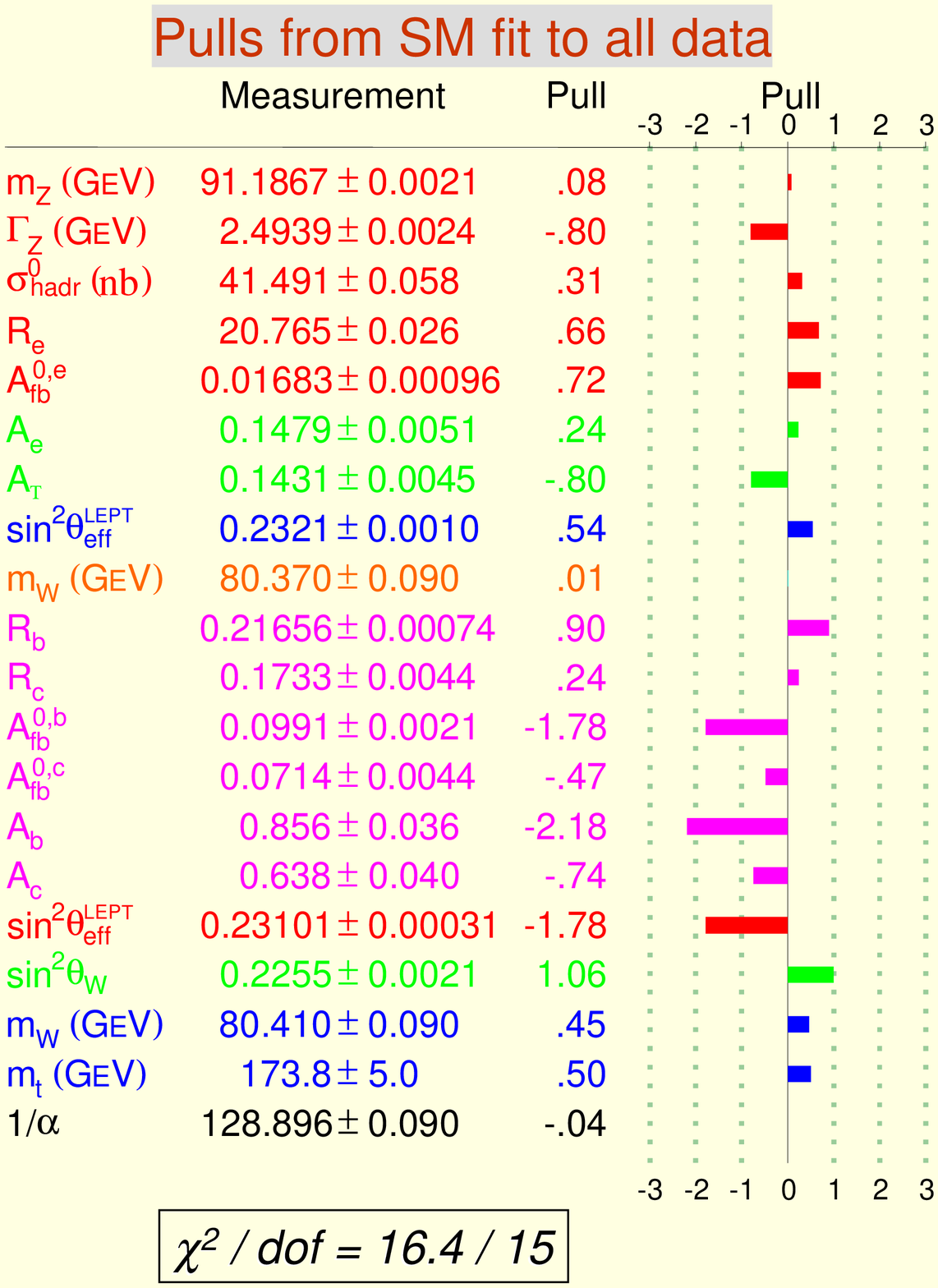}
\end{center}
\end{figure}

This fit gives 
$$ m_t = 171.6 \pm 4.9 \; Gev$$

$$m_H = 159.1^{+144}_{-83} \; Gev$$

$$\hat{\alpha}_s (m_z) = 0.119 \pm 0.003 $$

$$\chi^2/n.d.f. = 16.4/15$$
The quality of this fit is very good. We conclude that the SM gives the perfect description of $Z$ physics. New physics can not improve the fit of LEP and SLC data. Thus the Standard Model has been confirmed up to the loop corrections. What is more important is that  the loop corrections can be used to gather data on the not yet discovered particle. For instance, even before $t$-quark was discovered at Tevatron, its mass was predicted by analyzing the loops and LEP-SLC data. The hunting for virtual top quark is a very bright example of the collaboration of the theory with the experiment.  

\vspace{5mm}

{\bf 5.1. Decoupling of heavy flavors from Low-Energy} \\
\hspace*{9mm} {\bf Physics in QED and QCD.}

\vspace{3mm}

It is interesting to understand why in 1950's nobody worried about the contribution of top quark (and other heavy flavours) into magnetic moment of the electron known with very high accuracy. The answer to this question is that for $q \sim m_e$ the corrections due to top quark are suppressed as a power of 
$(m_e^2/m_t^2)$ i.e. the contribution was negligible. In QED we have decoupling of heavy (unknown) particles from the low-energy observables. Why? 
Consider the contribution of $t$-quark into QED observables. The only diagram with $t$-quarks in loop is the self-energy of the photon

$$
\Pi_{\mu\nu} (q) = i \langle 0| \{j_{\mu} (q),
 j_{\nu} (-q)\} |0 \rangle \eqno(5.1) 
$$
where $j_{\mu} (q)$ is the electromagnetic current of $t$-quark. Self energy has dimension 2: $[\Pi_{\mu\nu}] = m^2$. So one can expect that there exit terms of the order of 

$$ \Pi_{\mu\nu} \sim \alpha m_t^2 g_{\mu\nu}$$
This expectation is wrong in the case of \underline{conserved}
currents

$$ q_{\mu} j_{\mu} (q) = 0 \eqno(5.2)$$
Indeed for conserved current the self-energy operator should be transversal $q_{\mu} \Pi_{\mu\nu} = 0$. So 

$$ \Pi_{\mu\nu} (q) = (q_{\mu\nu}q^2 - q_{\mu}q_{\nu}) 
\Pi (q^2), \eqno(5.3)$$
Equation (5.3) implies that the photon remains massless. The scalar function $\Pi(q^2)$ has dimension zero and the only possible contribution of $t$-quark into $\Pi(q^2)$ can be written in the form

$$ \Pi(q^2) \sim \alpha \ln \frac{\Lambda^2}{m_t^2 + q^2}$$
where $\Lambda$ is cut-off. The self-energy keeps the memory of heavy flavours! 

The crucial step is renormalization. For instance consider the Coulomb scattering. If we take into account the infinite chain of self-energy contribution into the photon propagator we get for Coulomb amplitude 

$$ T_{Coulomb} = \frac{e^2_0 (\Lambda)}{q^2 (1 
+ \Pi(q^2))} \eqno(5.4)$$

At low $q^2$ we reproduce the Coulomb-low

$$ T = \frac{e^2_{phys}}{q^2}  \eqno(5.5)$$
with 

$$ e^2_{phys} = \frac{e^2_0 (\Lambda)}{1 + \Pi(0)} \eqno(5.6)$$
When we rewrite the amplitude (5.4) in terms of $e^2_{phys}$ we get 

$$ T \simeq \frac{e^2_{phys}}{q^2 [1 + 
\Pi(q^2) - \Pi(0)]} \eqno(5.7).$$
As a result: \\
1) the dependence on cut-off $\Lambda$ disappears 

$$ \Delta \Pi = \Pi (q^2) - \Pi (0) \sim \alpha \ln 
\frac{m_t^2}{m_t^2 + Q^2};$$
2) the contribution of heavy flavour is suppressed as a power 
$(q^2/m_t^2)$:

$$ \Delta \Pi \sim - \alpha  \left( \frac{q^2}{m_t^2} \right)
\rightarrow 0.$$
This is the sample of the famous decoupling theorem. It works for the theories with conserved vector currents.

\vspace{5mm}
{\bf 5.2. Non-decoupling of chiral matter.} \\
\hspace*{9mm} {\bf Heavy Flavor contribution into electroweak observables.}
\vspace{3mm}

In the Standard Model the left components of $t$- and $b$-quarks belong to $SU(2)_W$ doublet representation:
 $Q_L = \left( t_L \atop b_L \right)$. Therefore for the case when 
$m_t \gg m_b$ and for small energies $E \leq m_t$ we have effectively the  explicit violation of $SU(2)_W$ symmetry. 
For the virtual momenta $q \sim \Lambda \sim m_t$ theory looks like the old nonrenormalizable theory. It mean that one-loop corrections diverge quadratically 
$\delta_1 \sim \alpha \Lambda^2/m^2_Z$, two-loop corrections diverge quartically 
$\delta_2 \sim \left( \alpha \Lambda^2/m^2_Z \right)^2$. So we expect that the corrections to the low-energy observables due to top contribution are of the order of 

$$ \delta_1 \sim \alpha_W t$$

$$ \delta_2 \sim \alpha^2_W t^2$$
where $t = m_t^2/m^2_Z$, i.e.corrections are not suppressed, they grow  with top mass $m_t$. Heavy flavours are not decoupled from the low-energy observables for the chiral matter. As a result the radiative corrections in the SM are sensitive to the top contribution.

Consider for example the ratio of $g_V$ and $g_A$ for the decay 
$Z \rightarrow \bar{l}l$. It is rather simple exercise to calculate 
this ratio 

$$ R_l = \left( \frac{g_V}{g_A} \right)_l = 
1 - 4s^2 - \frac{3 \bar{\alpha}}{4\pi (c^2 - s^2)} (t + \delta V_R) \eqno(5.8)$$

The linear term $\sim t$ originates from different self-energies with top quark in the loops and $\delta V_R$ is the contribution of the rest of the diagrams (a few dozens ones) that do not depend on $t$. 

If we compare experimental value of $t$ 

$$ t^{\exp} \simeq 3.7$$
with experimental value of $(t + \delta V_R)$ 

$$ (t + \delta V_R)^{\exp} \simeq -0.49 \pm 0.32$$ 
we find that the constant terms are as much important as a linear term.

So we have to perform the accurate calculation of the whole set of one-loop diagrams.

\vspace{5mm}
{\bf 5.3. Radiative corrections in SM.}
\vspace{3mm}

a) \underline{Strategy.} \\
There are 3 steps in the calculations of the radiative corrections.

1) Calculate all observables in terms of base parameters $g_1^B$, 
$g_2^B$, $\eta^B$ and cut-off $\Lambda$. 

2) Find 3 \underline{basic observables} and express base parameters in terms of these 3 physical observables and cut-off $\Lambda$.

3) Substitute these expressions into formulas for other observables. Dependence on cut-off \underline{disappears}.
 
\vspace{3mm}

b) \underline{Basic parameters.}

It is reasonable to calculate observables in terms of quantities that
are known to the highest degree of precision.

$$ \alpha^{-1} = 137.035985 (61)$$

$$ G_{\mu} = 1.16639 (2) \cdot 10^{-5}  Gev^{-2} \eqno(5.9)$$

$$m_Z = 91.1867 (21) Gev$$

c) \underline{The Choice of Born approximation}
\vspace{3mm}

In contrast to $\alpha (q^2)$ that is \underline{running} 
coupling constant, the two electroweak coupling constants 
$\alpha_Z (q^2)$ and $\alpha_W (q^2)$ are not "running" but "crawling" for $q^2 \leq m^2_Z$. 

The natural scale for electroweak physics is $q^2 = m^2_Z$.
 Therefore it is evident that $\bar{\alpha} \equiv \alpha (m^2_Z)$, not $\alpha \equiv \alpha (0)$
is relevant parameter in electroweak physics. The value of 
$\bar{\alpha}$ is less accurate.

$$\bar{\alpha} = \alpha (m^2_Z) = [128.896 \pm 0.090]^{-1} \eqno(5.10)$$

With this parameterization it is convenient to introduce the weak angle $\theta$ by the relation 

$$ G_{\mu} = \frac{\pi \bar{\alpha}}{\sqrt{2} m^2_Z s^2 c^2} \eqno(5.11)$$
where $s^2 = \sin^2 \theta$, $c^2 = \cos^2 \theta$. Numerically 

$$ s^2 = 0.2311 (22) \eqno(5.12)$$  

\vspace{3mm}

d) \underline{Basic relations for electroweak observables.}

A simple calculation gives the following result for one-loop electroweak corrections in the case gluon free observables:

$$ \frac{m_W}{m_Z} = c \left[ 1 + \frac{3 \bar{\alpha}}{32 \pi s^2 (c^2 - s^2)} (t + \delta V_m) \right] $$

$$ g_{Al} = - \frac{1}{2} - \frac{3\bar{\alpha}}{64 \pi s^2 c^2}
(t + \delta V_A) \eqno(5.13)$$

$$ R_l = (g_V/g_A)_l = 1 - 4s^2 + \frac{3\bar{\alpha}}{4 \pi (c^2 - s^2)}
 (t + \delta V_R) $$
There are a few diagrams that contribute into leading term $\sim t$ and 
there are dozens of them for nonleading $\delta V$.

The hadronic decays of $Z$ are more sensitive to the value of the
gluonic coupling constant $\alpha_s$.

$$ \Gamma (Z \rightarrow q \bar{q}) = \frac{G_{\mu} m^3_Z}{2\sqrt{2}\pi}
[g_A^2 R_A + g^2_V R_V] \eqno(5.14)$$

The "radiator"  $R_{A,V}$ contain QCD and QED corrections caused by the final state emission and exchange of gluons and photons. The are hundreds diagrams that contribute into $R_{A,V}$.

\vspace{5mm}
{\bf 5.4. Hunting for virtual top. Great success.}

\vspace{3mm}

A comparison of the experimental data with the result of theoretical calculation eq.(5.13) led to the prediction of the $t$-quark mass $m_t$. 

$$ m_t = 180 (5)^{+17}_{-20} \quad  Gev$$
where the number in parentheses is the uncertainty due the uncertainties of the data. The center value corresponds to the assumption that 
$m_H = 300$ Gev, the upper and lower "shifts" correspond to 
$m_H = 1000$ Gev and $60$ Gev, respectively. 

The best fit of all observables gives 

$$ (m_t) \simeq 173.8 \pm 5.3 \quad Gev$$

These numbers are in perfect agreement with the recent direct measurement of the top-quark mass by two collaboration at FNAL 

$$ m_t = 173.8 \pm 5.0 \quad Gev.$$

\vspace{5mm}
{\bf 5.5. Hopeless hunting for virtual Higgs.}
\vspace{3mm}

Consider the limit of very large Higgs boson mass $m_H$. 
For $E \ll m_H$ we have $SU(2)$ symmetric theory of massive gauge bosons, i.e. effectively nonrenormalizable theory. As I explained in the 
Lecture II due to the gauge symmetry the leading divergence of the loop disappears. So the one-loop corrections diverge logarithmically 

$$ \delta_1 \sim \alpha_W \ln \frac{\Lambda^2}{m^2_Z} \sim \alpha_W 
\ln \frac{m^2_H}{m^2_Z}  = \alpha_W \ln h$$ 

two-loop corrections diverge quadratically

$$ \delta_2 \sim \alpha_W^2 \left( \frac{\Lambda^2}{m^2_Z} \right) 
\sim \alpha^2_W h.$$
Here $h = m_H^2/m_Z^2$.

This is the famous Veltman screening theorem.
The weak dependence of radiative corrections on $h$ results in a rather poor accuracy for $m_H$ derived from the precision data. The central value of $m_H$ from the fit should not be taken seriously. 
It is very unstable. Any tiny corrections or any change of the parameter can shift it by the order of magnitude. 

The one sigma upper bound is more reliable. According to the recent fit 

$$ m_H < 280 Gev~~~~~~~~~~~~95\%  \; c.l.$$

It seems that the fit of the precision data is not the best way for hunting for Higgs boson.
\newpage
  \centerline{REFERENCES}

\begin{description}         
\item [1]. Steven Weinberg, 
           "The Quantum Theory of Fields:Foundations", Vol.1,                      
            Cambridge Univ.Pr.,1995  and
           "Quantum Theory of Fields:Modern Applications",Vol.2,
            Cambridge Univ.Pr.,1996 .                    
         
            Michael E.Peskin, Daniel V.Schroeder, 
           "An Introduction to Quantum Field Theory",
            Addison-Wesley Pub.Co.,1995

            Martinus Veltman,
           "Diagrammatica:The Path to Feynman Rules",
            Cambridge Univ.Pr.,1994
         
            Lev B.Okun, "Leptons and Quarks" ,Amsterdam,North-Holland,1982

\item [2]. 't Hooft G., Nucl. Phys. {\bf B33}, 173 (1971), {\bf 35},
167 (1971).

 Veltman M., Nucl. Phys. {\bf B123}, 89 (1977); Acta Phys.
Pol. {\bf B8}, 475 (1977); \\

Passarino G., Veltman M., Nucl. Phys. {\bf B160}, 151 (1979).\\

 Berman S.M., Sirlin A., Ann. Phys. (N.Y.) {\bf 20}, 20
(1962); \\

Sirlin A., Rev. Mod. Phys. {\bf 50}, 573 (1978).

Sirlin A., Phys. Rev. {\bf D22}, 971 (1980); \\

Marchiano W.J. and Sirlin A., Phys. Rev. {\bf D22}, 2695 (1980); \\

Degrassi G., Fanchiotti S. and Sirlin A., Nucl. Phys. {\bf B351},
49 (1991).

 Fleischer J. and Jegerlehner F., Phys. Rev. {\bf D23},
2001 (1981).

 Aoki K.I., Hioki Z., Kawabe R., Konuma M. and Muta T.,
Suppl. Prog. Theor. Phys. {\bf 73}, 1 (1982).

  Consoli M., LoPresti S. and Maiani L., Nucl. Phys.
{\bf B223}, 474 (1983); \\

Barbieri R., Maiani L., Nucl. Phys. {\bf B224}, 32 (1983).

 Lynn B.W., Peskin M.E., Report SLAC-PUB-3724 (1985)
(unpublished); \\

Lynn B.W., Peskin M.E., Stuart R.G.,
{\it in} Physics at LEP (Report 
CERN 86-02) (CERN, Geneva, 1986), p. 90. 

 Kennedy D.C. and Lynn B.W., Nucl. Phys. {\bf B322}, 1
(1989).

 Bardin D.Yu., Christova P.Ch. and Fedorenko O.M.,
Nucl. Phys. {\bf B175}, 235 (1980); \\

Bardin D.Yu., Christova P.Ch. and Fedorenko O.M.,
Nucl. Phys. {\bf B197}, 1 (1982); \\

Bardin D.Yu., Bilenky M.S., Mitselmakher G.V., Riemann T. and

Sachwitz M., Z. Phys. {\bf C44}, 493 (1989).

 Bardin D. et al., Program ZFITTER 4.9, Nucl. Phys.
{\bf B351}, 1 (1991); Z. Phys. {\bf C44}, 493 (1989); Phys. Lett.
{\bf B255}, 290 (1991); Preprint CERN-TH.6443-92 (1992).

Altarelli G., Barbieri R., Phys. Lett. {\bf B253},
161 (1991); \\

Altarelli G., Barbieri R. and Jadach S., Nucl. Phys. {\bf B369}, 3
(1992), Erratum-ibid {\bf B376}, 446 (1992); \\

Altarelli G., Barbieri R., Caravaglios F., Nucl.Phys. {\bf B405},
3 (1993),
Phys. Lett. {\bf B314}, 357 (1993), Phys. Lett. {\bf B349}, 145
(1995).

 Ellis J., Fogli G., Phys. Lett. {\bf B213}, 189, 526
(1988); {\bf 232}, 139 (1989); {\bf 249}, 543 (1990).

 Hollik W., Fortschr. Phys. {\bf 38}, 3, 165 (1990); \\

Consoli M., Hollik W., Jegerlehner F., Proc. of the Workshop on $Z$
physics at LEPI (CERN Report 89-08) Vol. I, p. 7; \\

Burgers G. et al., ibid. p. 55.

 Montagna G. et al., Nucl. Phys. {\bf B401}, 3 (1993); \\

Montagna G. et al., Program TOPAZO, Comput. Phys. Commun. {\bf 76},
328 (1993).

 Chetyrkin K.G., K\"{u}hn J.H., Phys. Lett. {\bf B248},
359 (1990); \\

Chetyrkin K.G., K\"{u}hn J.H., Kwiatkowski A., Phys. Lett. {\bf 282},
221 (1992).

 Gorishny S.G., Kataev A.L., Larin S.A., Phys. Lett.
{\bf B259}, 144 (1991); \\

Surguladze L.R., Samuel M.A., Phys. Rev. Lett. {\bf 66}, 560 (1991).

Novikov V., Okun L., Vysotsky M., Nucl. Phys. {\bf B397},
35 (1993); \\

Vysotsky M.I., Novikov V.A., Okun L.B., Rozanov A.N.,
Uspekhi Fiz. Nauk, v. 166, 539 (1996) (Russian), Physics-Uspekhi
{\bf 39}(5), 503 (1996) (English); \\

Novikov V.A., Okun L.B. and Vysotsky M.I., Mod. Phys. Lett. {\bf A8},
2529 (1993); 3301 (E).

\item[3].  Reports on the working groups on precision
calculations for the $Z$ resonance (Report CERN 95-03), (CERN,
Geneva, 1995), p. 7-163.
\end{description}

\end{document}